\documentclass[final,5p,times,twocolumn]{elsarticle}
\usepackage{mypreamble}
\usepackage{amsmath,amsfonts}
\usepackage{algorithmic}
\usepackage{array}
\usepackage{braket}
\usepackage{textcomp}
\usepackage{stfloats}
\usepackage{url}
\usepackage{verbatim}
\usepackage{graphicx}
\usepackage{bm}
\usepackage[retainorgcmds]{IEEEtrantools}

\begin{document}

\begin{frontmatter}

\title{Audio compression using Periodic Gabor with Biorthogonal Exchange: \linebreak
Implementation Using the Zak Transform}

\author[mymainaddress,mysecondaryaddress]{Roger Alimi \corref{mycorrespondingauthor}}
\cortext[mycorrespondingauthor]{Roger Alimi}
\ead{roger@soreq.gov.il}

\author[mysecondaryaddress]{David J. Tannor}

\address[mymainaddress]{Technologies Division at the Soreq Nuclear Research Center, Yavneh 81800, ISRAEL}
\address[mysecondaryaddress]{Department of Chemical Physics, Weizmann Institute of Science, Rehovot, 76100, ISRAEL}

\begin{abstract}
An efficient new approach to signal compression is presented based of a novel variation on the Gabor basis set. Following earlier work by Shimshovitz and Tannor\cite{Shim1, Shim3}, we convolve the conventional Gabor functions with Dirichlet functions to obtain a Periodic Gabor basis set (PG). The PG basis is exact for continuous functions that are periodic band-limited. Using the orthonormality of the Dirichlet functions, the calculation of the PG coefficients becomes trivial and numerically stable, but its representation does not allow compression. Large compression factors are achieved by exchanging the PG basis with its biorthogonal basis, thereby using the localized PG basis to calculate the coefficients (PGB). Here we implement the PGB formalism using the Fast Zak Transform and obtain very high efficiency with respect to both CPU and memory. We compare the method with the state of the art Short-Time Fourier Transform (STFT) and Discrete Wavelet Transform (DWT) methods on a variety of audio files, including music and speech samples. In all cases tested our scheme surpasses the STFT by far and in most cases outperforms DWT.
\end{abstract}


\end{frontmatter}

\section{Introduction}

Building on an idea of von Neumann in quantum mechanics \cite{VonNeumann}, Gabor introduced a time-frequency representation of signals where the signal is expanded using a lattice of Gaussian basis functions \cite{Gabor}. These functions have the form:
\begin{equation}
\label{gnm}
g_{nm} = g(t-na)e^{imbt}.
\end{equation}
In this expression, $a$ and $b$ represent time and frequency intervals respectively, while $n$ and $m$ are integers. The function $g$ is a localized window shifted in time and frequency. Gabor suggested using these functions as a basis set in order to represent a continuous signal $s (t)$ by:
\begin{equation}
\label{st}
s(t) = \sum_{n=-\infty}^{\infty} \sum_{m=-\infty}^{\infty} c_{nm}g_{nm}(t).
\end{equation}
where $c_{nm}$ are the coefficients of the expansion. For an arbitrary signal, ({\ref{st}) has a solution only if $ab \le 2\pi$. For the case $ab = 2\pi$, called critical sampling, the Gabor lattice is complete on the infinite plane. The appeal of this basis is that it allows for time-frequency correlation, and therefore holds out the promise for a representation significantly sparser than e.g. a Fourier representation. However, the Gabor basis has presented three types of challenges.

First, although the Gabor lattice is complete on the infinite plane, it turns out that it is not complete on any finite subspace of the infinite plane. In fact, on a finite subspace the representation is unstable \cite{Daub}, meaning that the convergence goes as an inverse power of the number of basis functions, as opposed to exponentially. This instability arises from the non-orthogonality of the Gabor basis, i.e. that the $g_{nm}(t)$ functions cannot be made orthonormal without sacrificing the locality of $g(t)$ in either time or frequency \cite{Balian,Low}. The instability has been noted in the signal processing community as well as in the quantum mechanics community, where the Gabor representation goes by the name of the von Neumann lattice \cite{Davis,Poir}. 

Second, despite its intuitive promise, the method actually leads to a highly non-sparse representation, meaning there are surprisingly few negligible elements in the joint time-frequency representation (this is distinct from the instability of the method described above which refers to poor convergence in $t$ or $\omega$). The origin of the non-sparsity of the Gabor coefficients will be explained below; it has gone unmentioned or unexplained in many studies \cite{Daug,Porat,Ebra} but appears to be ubiquitous \cite{Bast5}.

Third, working with a non-orthogonal basis involves technical difficulties and additional formalism not familiar from orthogonal bases.

As a result of these difficulties, a different localized time-frequency expansion was developed, the “wavelet” expansion \cite{Daub}, in which time-frequency localization is achieved by shifting and scaling the synthesis function. In this approach one constructs a localized and orthogonal set of functions, bypassing the problems of the Gabor expansion. Wavelets are a form of time-frequency representation, but strictly speaking, in many implementations, they are not a basis but a tight frame.  They are considered state-of-the-art for signal or image compression \cite{Mall, Gro,Tse, Bar}.

Despite the success of wavelets, the Gabor approach has continued to attract attention. A major advance regarding the technical difficulties was made by Bastiaans \cite{Bast1} who showed that the coefficients of the Gabor functions can be represented as the inner product between the signal $s$ and a basis set $\gamma_{nm}$ that is biorthogonal to $g_{nm}$. Bastiaans’s approach does not solve either the stability problem or the sparsity problem: it just transforms the problem of finding $c_{nm}$ to the problem of finding $\gamma_{nm}$. For the few special cases where 
$\gamma_{nm}$ can be calculated analytically, the Gabor coefficients are indeed found to be non-sparse.  

In parallel, there has been significant progress on the stability problem. Wexler and Raz \cite{Wex} and Orr \cite{Orr1} considered a variant of the Gabor method based on periodization and sampling, analogous to the procedure for obtaining the finite Discrete Fourier Transform from the Fourier integral. If the conditions of Nyquist's theorem are satisfied, the finite discrete variant of the Gabor method allows for an exact reconstruction of the signal for continuous periodic, band-limited signals. Many signals that do not rigorously meet this criterion can still be approximated in this way if they converge exponentially in time and frequency.  For instance, the finite Discrete Fourier Transform and its implementation as the Fast Fourier Transform allow exact reconstruction for precisely this same class of signals and are of enormous utility even when signals do not precisely meet these criteria. Thus, the discrete, periodic Gabor representation solves the stability problem with the Gabor representation for an important class of signals. However, as far as we know no publication has explicitly pointed out the connection between the finite, discrete Gabor formalism and the solution to the unstable behavior of the continuous Gabor representation.

Shimshovitz and Tannor (ST) \cite{Shim1} came to the solution of the stability problem of the Gabor basis from a different route. They asked how one could make the time-frequency (TF) coverage of the Gabor basis identical to that of a band-limited (BL) Fourier series. By considering the BL Fourier series in time-frequency space, it became apparent that at critical sampling the Gabor basis spanned the same TF space with the same number of basis functions, but had different boundary conditions. By convolving the conventional Gabor functions with Dirichlet (periodic sinc) functions, they obtained a Periodic Gabor basis having the same boundary conditions as the BL Fourier series.  The PG basis is therefore related by a unitary transformation to the BL Fourier series and allows for exact reconstruction of periodic band-limited signals. 

With regard to the non-sparsity of the Gabor representation there has been less progress. In some cases, the phenomenon has been addressed by oversampling i.e. $ab > 2\pi$ \cite{Wex,Qian,Boo}: when the oversampling is increased sufficiently, the biorthogonal set $\gamma_{nm}$  become similar to the Gabor set $g_{nm}$, and the coefficients become sparse. However, oversampling is inefficient in the sense that a time-frequency region that was previously overlapped by one Gabor function is now overlapped by many, and all of them need to be taken into account in the expansion. An innovative way to overcome the non-sparsity without oversampling was developed by ST. They showed that the origin of the non-sparsity is the delocalized character of the biorthogonal basis. By exchanging the roles of the PG basis and its biorthogonal basis, i.e. using the localized PG basis to calculate the coefficients rather than as the basis, they were able to achieve extremely high sparsity. They called the method Periodic Gabor with Biorthogonal exchange or PGB. Like the PG basis, the method is related to the BL Fourier series by a unitary transformation and gives the same numerical results to machine precision if no compression is performed.

To reiterate, both the PG and PGB methods are related to the BL Fourier series by a unitary transformation; as such both are bona fide bases, not frames. However, the PG basis allows for almost no compression, while the PGB method allows for significantly larger compression factors than does the BL Fourier series, and therefore the rest of this article will deal exclusively with the PGB method and its extension. The fact that the PGB method is a bona fide basis distinguishes it from many of the time-frequency representations in common use, such as Short Time Fourier Transform (STFT) or many implementations of the Discrete Wavelet Transform (DWT).

Although we have presented the work of ST in terms of the variables $\omega$-$t$, almost all their work was in the context of quantum mechanics, where the conjugate variables are momentum $p$ and position $q$ instead of $\omega$ and $t$. The lattice of $p$-$q$ Gaussians is referred to as the von Neumann lattice (note that von Neumann's work \cite{VonNeumann} preceded that of Gabor \cite{Gabor} by fifteen years). Thus, ST referred to their method as “Periodic von Neumann with Biorthogonal exchange”, or PvB. In \cite{Mach ,Shim2 ,Take}, the PvB was applied to quantum mechanics with great success. ST also applied the method to signal processing where, as mentioned above, they called it ``Periodic Gabor with Biorthogonal exchange" or PGB. Promising results were obtained for both audio and image processing \cite{Shim3}, but the method was expensive in both CPU and memory due to the need to invert a large matrix; although this work was posted on the archives it was never published in the journal literature.

In this paper:
\begin{itemize}
\item {We show that the expensive CPU and memory in \cite{Shim3} can be avoided by combining PGB with the Fast Zak Transform. The Fast Zak Transform scales as $N \log N$, the same as the FFT. However, even with the Fast Zak Transform, PGB would still require the inversion of a large matrix to find the biorthogonal basis. We overcome this problem by exploiting a property of the Zak Transform of biorthogonal functions that completely avoids the need for matrix inversion.}
\item {The Fast Zak Transform allows many orders of magnitude savings both in CPU time and memory over the previous PGB method. In \cite{Shim3}, the audio segments were limited to about 1s; in contrast, the combined method of PGB with Fast Zak Transform (PGBZ) can handle audio segments of several minutes. We apply the method to a wide range of audio signals, including musical instruments, speech and finally even to a 2.5-minute segment of Beethoven's 6th symphony.}
\item {Systematic comparisons are performed between PGBZ, the Short-Time Fourier Transform (STFT) and the Discrete Wavelet Transform (DWT) methods. All three methods have the same overall scaling of $N \log N$, but PGBZ allows more than an order of magnitude more compression than STFT and is competitive with, and generally superior to, DWT.}
\item {A downside of the PGBZ compression is a small, residual periodic noise after the compression. However, this is easily removed using commercial noise filtering software; in fact, we show that this noise filtering is mimicking the rigorous but computationally expensive Porat reorthogonalization procedure for the biorthogonal basis.}
\end{itemize} 

A note about terminology. In this paper we use the term `compression' for finding a representation in which the signal is sparse and then removing coefficients with low amplitude. This reduces the computational requirements for storing and reconstructing the signal, at the expense of introducing some error in the reconstructed signal.  For commercial compression, entropy coding and quantization are normally employed subsequent to this sparsification process, but the sparsification we do in this paper is the first step, and in and of itself falls under the rubric of the standard definition of lossy compression.  

The organization of the paper is as follows. In Section II we review the theory of the Periodic Gabor method with Biorthogonal exchange. In Section III we describe the efficient implementation of the method using the Zak transform. Section IV presents numerical results, including comparisons with STFT and Wavelet methods and describes the noise filtering procedure. Section V concludes.

\section{The Periodic Gabor Biorthogonal Representation}

Equations (1) and (2) define the continuous and infinite Gabor representation. A finite, discrete variation can be formulated by analogy with the method used to obtain the finite Discrete Fourier Transform from the Fourier integral (see for instance Wexler and Raz \cite{Wex} and Orr \cite{Orr1}). If the conditions of Nyquist's theorem are satisfied, the discretized variant of the Gabor method allows for an exact reconstruction of the signal for continuous, periodic, band-limited signals. Many signals that do not rigorously meet this criterion can still be approximated in this way if they converge exponentially in time and frequency; the Discrete Fourier Transform and its implementation as the Fast Fourier Transform allow exact reconstruction for precisely this same class of signals and are of enormous utility even when signals do not precisely meet these criteria.

Shimshovitz and Tannor \cite{Shim1} came to periodization of the Gabor functions by a very different route.  They 
first noted that the Gabor basis spans the same time-frequency space as a band-limited Fourier series with the same number of functions. Consider a time interval $[0,T]$ and a frequency interval $\omega_{\mathrm {range}} \equiv \omega_{\mathrm {max}} - \omega_{\mathrm {min}}$ such that the total TF area is 
\begin{equation}
T \times \omega_{\mathrm {range}} \equiv 2 \pi N,
\label{eq:2piN}
\end{equation}
where $N$ is defined by eq. \ref{eq:2piN}. Consider first a set of $N$ Gabor functions with $N_{t}$ cells in time and $N_{\omega}$ in frequency such that $N_{t}N_{\omega} = N$. Since each Gabor basis functions spans a TF space of $2\pi$, the truncated Gabor basis spans a TF area $2\pi N$. Now consider the BL Fourier series. If the time interval is $[0,T]$, periodization implies $\delta \omega=2 \pi/T$. The number of Fourier functions necessary is then 
\begin{equation}
\frac{\omega_{\mathrm {range}}}{\delta \omega}= \frac{2 \pi N/T}{2 \pi/T} = N,
\end{equation}
the same as the number of Gabor functions.

\begin{figure}[!t]
\centering
\includegraphics[width=4in]{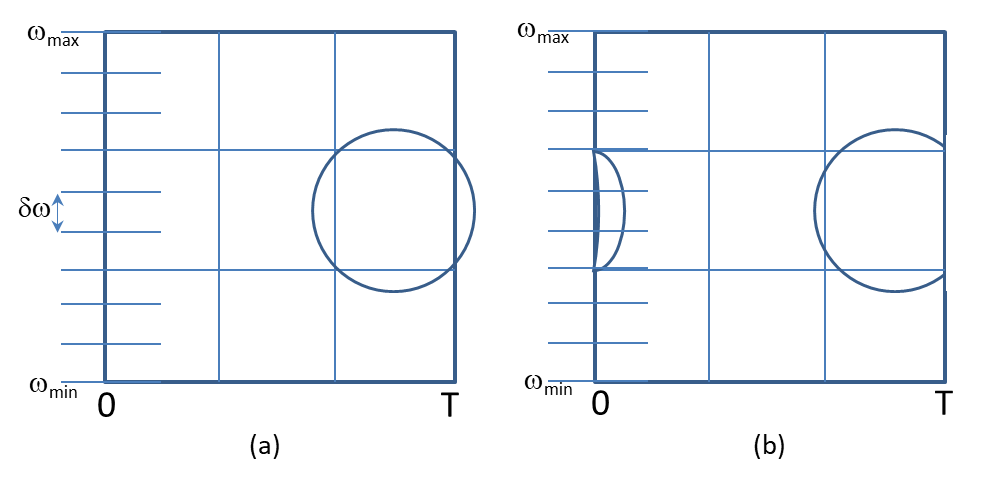}
\caption{(a) 9 Gabor unit cells and 9 values of a BL Fourier series cover the same area in TF space, $S=2\pi N$. Superimposed is a schematic Gabor function. (b) The periodic Gabor basis has the same boundary conditions as the BL Fourier series and is therefore a complete set for the truncated space.}
\label{fig1}
\end{figure}
However, the Gabor functions protrude beyond the boundaries of the TF space (Fig. 1a); since they partially cover the exterior region, this hints that they are not complete on the interior. In contrast, the BL Fourier series truncates sharply at 
$\left [ \omega_{\mathrm {min}},\omega_{\mathrm {max}}\right] \otimes \left [ 0 \text{ and } T \right]$ (Fig. 1b). The key idea in the derivation of Shimshovitz and Tannor is to modify the boundary conditions of the TF space spanned by the truncated Gabor basis so that they are identical with those of a BL Fourier series. Since the BL Fourier series is complete on the space of band limited periodic functions, the modified Gabor basis will also be complete on that space.

To make these ideas precise, Shimshovitz and Tannor formulated both approaches in a way that facilitates direct comparison. They combined the Gaussian and the Fourier basis functions to generate a ``Gaussian-like" basis set that is confined to the truncated space. For this purpose they used the basis sets $ \{ g_{nm}(t) \} $ and $ \{ \theta_{i}(t) \} $ to construct a new basis set, $ \{ \tilde{g}_{nm}(t) \} $:
\begin{equation}
\label{gnmt}
\tilde{g}_{nm}(t) = \sum_{i=1}^{N} \theta_{i}(t) g_{nm}(t_{i}),
\end{equation}
for $n=1,...,N_{t},~m=1,...,N_{\omega}$ and $t_i,~i=1,...N$ are discrete values of time spaced by the Nyquist spacing, $\delta=T/N$. The basis functions  $ \{ \theta_{i}(t) \} $ are given by (\cite{Tann1}, eq. 11.163):
\begin{IEEEeqnarray}{rCl}
\theta_{i}(t) &=& \sum_{j=(-N/2) +1}^{N/2} \frac {1}{\sqrt{TN}} \exp \left [ \frac {2 \pi \sqrt{-1} j} {T} (t-t_{i}) \right ]  \\ 
&=& \frac {e^{i\pi(t-t_i)/T}} {\sqrt{TN}} \frac{ \sin{ \left [ \pi(t-t_i)N/T \right] }  }{\sin{ \left [ \pi (t-t_i)/T \right ] } },
\label{thetak}
\end{IEEEeqnarray}
which up to a phase factor are periodic sinc functions or Dirichlet functions $D_N(t)$ \cite{matdir} \cite{Tann1}. The Dirichlet functions for different indices $i$ are orthonormal (\cite{Tann1}, eq. 11.163 with eq. 11.47), a property that we will use below. The new basis set is, in some sense, the Gaussian functions, modified to be band-limited with periodic boundary conditions. Unlike the original Gaussian basis $\bf{G}$,
the new basis $\bf{\tilde{G}}$ spans the same Hilbert space as the BL Fourier series. This property provides us with an accurate and stable time-frequency representation. 

The signal $s(t)$ is now expanded as:
\begin{equation}
\label{stsum}
s(t) = \sum_{k=1}^{N}  c_{k} \tilde {g}_{k}(t).
%\label{eq:signal_gabor}
\end{equation}
For convenience we adopt a single index notation for the Gabor functions except when otherwise specified. The single notation is mathematically equivalent to the double index notation as long as we carefully perform a suitable packing/unpacking of the relevant arrays; the packing is given by  $k = n+m+N_{t}(m-1)$, for $n=1,...,N_{t}$, $m = 1,..., N_{\omega}$ and $N = N_{t}N_{\omega}$. Note that the periodicity of the new basis $\tilde{\bold {G}} $ allows us to replace the infinite sum of (\ref{st}) by a finite number of terms in (\ref{gnmt}).

We now turn to the calculation of the coefficients $c_k$. Left multiplying both sides of (\ref{stsum}) by $ \tilde{g}_{j}^{*}(t)$ and integrating over $t$, we obtain:
\begin{equation}
\label{tildegj}
\braket{\tilde{g}_j,s} = \sum_{k=1}^{N} c_{k} \braket{\tilde{g}_j , \tilde{g}_k} = \sum_{k=1}^{N} c_{k} S_{jk} ,
\end{equation}
where we have defined the continuous inner product in a complex vector space as 
$\braket{ \tilde {f}_m , \tilde {g}_n} = \int_{0}^{T} \tilde {f}_m^*(t) \tilde {g}_n(t) dt$ and $\bold{S}$ is the 
%with the understanding that for band-limited periodic functions the continuous inner product is given exactly by the sum over the product values at the Nyquist sampling points. Note that the inner product is defined in a complex vector space, where the left element is assumed to be complex conjugated.
overlap matrix with elements $S_{jk} \equiv \braket{\tilde{g}_j , \tilde{g}_k}$.
Equation (\ref{tildegj}) can be written in matrix form as:
\begin{equation}
\label{bolds}
\bold {s} = \bold{Sc} .
\end{equation}
Left multiplying both sides by  $\mathbf { S^{-1}}$ we obtain $\mathbf{c = S^{-1}s}$, or
\begin{equation}
\label{ck}
c_k = \sum_{j=1}^{N} S_{kj}^{-1}s_j =  \sum_{j=1}^{N} S_{kj}^{-1} \braket{ \tilde{g}_j ,s}.
\end{equation}
Substituting (\ref{ck}) for $c_k$ back into (\ref{stsum}) we obtain
\begin{equation}
\label{stsumk}
s(t)  = \sum_{k=1}^{N}\sum_{j=1}^{N} \tilde{g}_{k}(t)S_{kj}^{-1} \braket{\tilde{g}_j,s}.
\end{equation}
It is convenient to define the new quantity
\begin{equation}
\label{gammak}
\tilde {\gamma}_k (t) \equiv \sum_{j=1}^{N}\tilde{g}_{j}(t)S_{jk}^{-1} 
\end{equation}
for $k = 1,...N$. Using the fact that $\bold{S}$ and therefore $\bold{S}^{-1}$ is Hermitian, $\tilde{\gamma}_k^*(t)= \sum_{j=1}^{N}S_{kj}^{-1} \tilde{g}_j^*(t)$. In terms of $\tilde {\gamma}_k$, (\ref{ck}) becomes
\begin{equation}
\label{ckbrakket}
c_k = \braket{ \tilde {\gamma}_k ,s},
\end{equation}
and (\ref{stsumk}) becomes:
\begin{equation}
\label{stsumk1}
s(t)  = \sum_{k=1}^{N}  \tilde{g}_k(t) \braket{ \tilde {\gamma}_k ,s}.
\end{equation}
It is easily checked that the $\tilde{\gamma}$ are biorthogonal to the $\tilde{g}$, that is:
\begin{equation}
\label{braketggamma}
\braket{ \tilde{g}_k , \tilde{\gamma}_l } = \braket{ \tilde{\gamma}_k , \tilde{g}_l }  = \delta_{kl} .
\end{equation}

Shimshovitz and Tannor noted that although the $\bold {S}$ matrix is localized (i.e. banded), 
$\bold {S}^{-1}$ is not.  This leads to a series of consequences. 
From (\ref{gammak}) this implies that although the $\tilde{g}$ are localized the $\tilde{\gamma}$ are delocalized. 
The delocalized character of the $\tilde{\gamma}$ in turn implies from (\ref{ckbrakket}) that virtually no $c_k$'s are small, i.e. the representation is non-sparse. Finally, since virtually all $c_k$ are non-negligible, this implies from (\ref{stsum}) that virtually all basis functions $\tilde{g}_k(t)$ contribute to the expansion of the signal $s(t)$, even those basis functions that are distant from the signal in time-frequency; this can be appreciated directly from (\ref{stsumk}) where even if only a few $\tilde{g}_j$ overlap the signal $s$ in time-frequency, the $S_{kj}^{-1}$ couples those few to essentially every $\tilde{g}_k$ and as a result virtually none of the $\tilde{g}_k(t)$ can be eliminated from the expansion.

In order to overcome the non-sparsity of the coefficients, Shimshovitz and Tannor noted that (\ref{stsumk}) can be rewritten as follows:
\begin{equation}
\label{stsumk2}
s(t)  = \sum_{k=1}^{N}  \tilde{\gamma}_k(t) \braket{ \tilde {g}_k ,s} \equiv \sum_{k=1}^{N}  a_k \tilde{\gamma}_k(t),
\end{equation}
where
\begin{equation}
a_k  = \braket{ \tilde {g}_k ,s},~~k = 1,...,N.
\label{eq:a_k}
\end{equation}
This exchanges the role of the Gabor basis and its biorthogonal basis, so that the delocalized $\tilde{\gamma}$ 
functions become the basis functions and the localized $\tilde{g}$ 
functions determine the coefficients. Although the 
$\left \{ \tilde{\gamma} \right\}$ 
basis is highly counterintuitive because of its delocalized character, the localized  $\tilde{g}$  
functions are used to calculate the coefficients, and therefore many of the $a_k$ 
will be close to zero. As a result, according to (\ref{stsumk2}) many of the  $\tilde{\gamma}$ 
can be eliminated from the expansion of $s(t)$, allowing for significant compression.
An example of a spectrogram arising in the PGB method is shown in the Supplementary Material.

The existence of the basis biorthogonal to the Gabor functions has been known in the signal processing field since the pioneering work of Bastiaans \cite{Bast2}.  However, the exchange of the roles of the Gabor and the biorthogonal basis (i.e. using $\left \{ \gamma\right\}$ as the basis and $\left\{g\right\}$ to calculate the coefficients) is new to the work of Shimshovitz and Tannor. In the next section we discuss how PGB still requires the inversion of a large matrix, and how this can be overcome using the Fast Zak Transform.

We close this section by noting a remarkable relationship between continuous functions and discrete sampling for functions that are band-limited and periodic: the continuous (Hilbert) inner product of two band-limited periodic functions,  $\tilde{f^{*}}(t)$ and $\tilde{g}(t)$ is reproduced exactly by the discrete inner product calculated using the values of the functions $f^{*}(t)$ and $g(t)$ (without the tildes!) at the Nyquist sampling rate.
%(weighted by $\sqrt{\delta t}$). 
To see this, substitute (\ref{gnmt}) for $\tilde{g}$ and the corresponding expression for $\tilde {f^{*}}$ into an expression for the continuous inner product (because we are dealing with general functions we omit the subscripts on $f(t)$ and $g(t)$):
%\begin{equation} \label{deqn_ex1}
\begin{IEEEeqnarray}{rCl}
\int_{0}^{T} \tilde { f}^{*}(t) \tilde{g}(t) dt&=&\int_{0}^{T} \sum_{i=1}^{N} \theta_{i}(t) f^{*}(t_{i}) \sum_{j=1}^{N} \theta_{j}(t) g(t_{j}) dt   \\
&=&  \sum_{i=1}^{N} f^{*}(t_{i}) g(t_{i}) .
\label{integfg}
\end{IEEEeqnarray}

In arriving at the last expression we have used the orthonormality of the Dirichlet functions:
\begin{equation}
\label{integ_theta}
\int_{0}^{T}  \theta_{i}(t)\theta_{j}(t) dt = \delta_{ij} .
\end{equation}
Equation (\ref{integfg}) can be viewed as a corollary of Nyquist’s theorem \cite{Tann2}. Just as Nyquist's theorem asserts that a continuous band-limited signal can be represented precisely in terms of its values at a discrete set of points with the Nyquist spacing, so too the inner product of two continuous band-limited signals can be represented precisely in terms of the product values at a discrete set of points with the Nyquist spacing. The continuous inner products in Eqs. (\ref{tildegj}) - (\ref{eq:a_k}) are therefore given exactly by the sum in Eq. (\ref{integfg}) for this class of functions.

\section{Computing PGB Coefficients Using The ZAK Transform}
The “straightforward” way to compute the PGB coefficients is to compute the full complex Gaussian basis, then the overlap matrix, invert the latter and finally multiply the inverse overlap  matrix by the original Periodic Gaussian basis. Assuming $N_t$ and $N_{\omega}$ Gaussians in $t$ and $\omega$ respectively such that $N_{t}N_{\omega} = N$, we have to compute a $ N \cdot N = N_{t}^{2}N_{\omega}^{2}  $ matrix, invert it, and multiply it by a column vector of size $N$.  Let us consider a small audio file of approximately 10 sec duration, sampled at 44100 Hz. Taking $N_t =  N_{\omega}$= 665, we get $N$ = 442225.  Computing, storing and inverting a complex matrix of 442225 $\times$ 442225 is practically impossible.

There are several ways to overcome the problem. First, one can divide the file into smaller time intervals. But even 1sec produces a matrix of almost 30Gb memory, still a lot to handle. Another option is to use sparse algebra, which is possible since almost all of the off-diagonal elements of the overlap matrix are very close to zero. This reduces the memory requirements by one order of magnitude but still requires working with files of small duration and Gigabyte size matrices. It also increases the CPU time and the code complexity. A much better solution, both from memory and CPU time considerations, exploits a different formalism, namely the Zak transform method \cite{Zak,Jans,Bast2}. 

The Zak transform of a function $f(t)$ is a mixed time-frequency mapping given by:
\begin{equation}
\label{zak1}
Z_{f}(t,\omega) = \sum_{k=-\infty}^{\infty} f(t+kT)e^{-2 \pi i k \omega T} , 0 \le t \le T, 0 \le \omega \le 1/T, 
%Z_{f}(t,\omega) = \sum_{-\infty}^{\infty} f(t+kT)e^{-2 \pi i k \omega b T} , 0 \le t \le T, 0 \le \omega \le 1/T
% WHAT ABOUT THE DIFFERENCE BETWEEN OMEGA AND NU?
\end{equation}
where $T$ is the sampling time.
The Zak transform divides an infinite time signal $f(t)$ into intervals $T$. The variable $t$ is then simultaneously swept  $0 \le t \le T$ within all the intervals; for each $t$ the discrete Fourier transform is calculated and plotted for $0 \le \omega \le 1/T$. In this way the Zak transform maps the infinite time signal to a rectangle $[0 \le t \le T$, $0 \le \omega \le 1/T]$.
The Zak transform has been widely used in electrical engineering, quantum field theory and signal and image processing 
\cite{AnJ16,ShL22,FrL16,Zh16,LaA21,Ab17}. 

We begin with a description of how the Zak transform is applied to the Periodic Gabor basis and then describe how it is modified for the biorthogonal exchange. 
By combining (\ref{gnm}) and the double index version of (\ref{stsum}) we obtain (at critical sampling $ab = 2\pi$):
\begin{equation}
\label{zak2}
s(t) = \sum_{n=1}^{N_t} \sum_{m=1}^{N_{\omega}} c_{nm} \tilde{g}(t-na)e^{2 \pi imt/a}.
%\label{eq:zak_signal}
\end{equation}
Taking the Zak transform of both sides of (\ref{zak2}) one gets:
\begin{multline}
\label{zak3}
Z_{s}(t,\omega)= \sum_{k=-\infty}^{\infty} \left [ \sum_{n=1}^{N_t} \sum_{m=1}^{N_{\omega}} c_{nm} \tilde{g}(t+kT-na) e^{2 \pi imt/a} \right ] \\ e^{-2 \pi i k \omega T}.
\end{multline}
The Zak transform is semi-periodic in the first argument \cite{Jans}, i.e.
\begin{equation}
\label{zak4}
Z(t+a,\omega) = e^{ 2 \pi i \omega a} Z(t,\omega).
\end{equation}
Rearranging (\ref{zak3}) and using the periodicity of the Zak transform, one can write:
\begin{IEEEeqnarray}{rCl}
\label{zak5}
Z_{s}(t,\omega)=  \left [ \sum_{n=1}^{N_t} \sum_{m=1}^{N_{\omega}} c_{nm} e^{2 \pi i (mt/a - n \omega a)}  \right] \nonumber \\ 
\cdot  \left [ \sum_{k= -\infty}^{\infty}  \tilde{g}(t+kT) e^{-2 \pi i k \omega T} \right].
\end{IEEEeqnarray}
The first factor of the right side of (\ref{zak5}) is the 2D Fourier transform of the Gabor coefficients, while the second factor is the Zak transform of the Periodic Gaussian basis. Therefore we have:
\begin{equation}
\label{zak6}
Z_{s}(t,\omega)=  \left [ \sum_{n=1}^{N_t} \sum_{m=1}^{N_{\omega}} c_{nm} 
e^{2 \pi i(mt/a - n \omega a)} \right ] \cdot Z_{\tilde g}(t,\omega).
%\label{eq:zak_coeff}
\end{equation}
Equation (\ref{zak6}) may be inverted to give:
\begin{equation}
\label{zak7}
c_{nm} = \int_{0}^{a} dt \int_{0}^{1/a} d\omega \left ( \frac {Z_{s}(t,\omega)}{ Z_{\tilde g}(t,\omega)} \right )  e^{2 \pi i (n \omega a -mt/a)},
\end{equation}
where the $c_{nm}$ are the Fourier coefficients of the ratio $Z_{s} /Z_{\tilde g}$.

Three comments about (\ref{zak7}):

\begin{enumerate}
\item Equation (\ref{zak7}) reveals a potential difficulty with the Zak transform: the ratio diverges if there are zeros in the Zak transform of the window function. Unfortunately, this is indeed the case for the Gaussian function as we can see in Fig. ~\ref{fig2}. It turns out that most of the time the divergence can be bypassed by shifting the center point of the map by half a grid point, which is equivalent to constraining the window size to be an odd number \cite{Assa,Ausl}.

\item Equation (\ref{zak7}) appears in Orr \cite{Orr2} and in Chinen and Reed \cite{Chin}. We have included the detailed derivation here for three reasons: firstly, for the sake of completeness; secondly, because as opposed to the case in\cite{Orr2} and \cite{Chin} the transition from the continuous to the discrete signal arises naturally (see (\ref{gnmt}) with (\ref{integfg}) ) rather than discreteness being imposed a priori. Thirdly, below we will need to modify the derivation to exchange the role of the primary and the biorthogonal basis to arrive at (\ref{zak8}).

\item In practice, the computation of the Gabor coefficients (or in our case, the Periodic Gabor coefficients) using the Zak transform involves three steps. a) Compute the Zak transforms of the signal and of the window function, b) take their ratio and finally c) perform a 2D inverse Fourier Transform of the result. The whole procedure may also be reversed: if a Zak-Gabor coefficient map is given, one can reconstruct the original signal by taking a 2D Fourier Transform of the map, multiplying by the Zak transform of the fiducial Gaussian and taking the inverse Zak transform of the result. 

\end{enumerate}

\begin{figure}[!t]
\centering
\includegraphics[width=3.8 in]{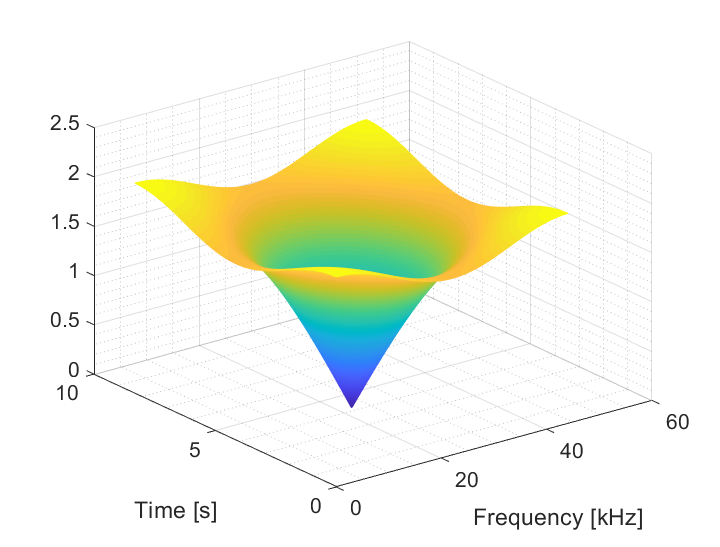}
\caption{The Zak transform (absolute value) of a Gaussian (8sec duration sampled at 44.1 kHz).}
\label{fig2}
\end{figure}
To implement the Zak transform in the biorthogonal exchange formalism requires just a small modification: we replace the fiducial Gaussian by its biorthogonal counterpart $\gamma$, leading to:
\begin{equation}
\label{zak8}
a_{nm} = \int_{0}^{a} dt \int_{0}^{1/a} d\omega \left ( \frac {Z_{s}(t,\omega)}{ Z_{\tilde \gamma }(t,\omega)} \right ) e^{2 \pi i (n \omega a -mt/a)}.
%\label{eq:a_nm}
\end{equation}
In other words the coefficients are the Fourier components of the ratio $Z_s/Z_{\tilde{\gamma}}$, which we refer to from now on as PGBZ coefficients  (Z for Zak). Again, we should take care to avoid zero values in the denominator. 

Equation (\ref{zak8}) provides a simple and straightforward method to compute the PGBZ coefficients. As stated by Chinen and Reed \cite{Chin}, this method is the only algorithm having the same order of computation as the DFT. However in order to compute the denominator we still have to compute the overlap matrix and its inverse, with the same heavy limitations previously described. But here is the trick: the Zak transform has an additional interesting property. In the continuation of his pioneering work in this field \cite{Bast3,Bast4}, Bastiaans proved that if $g$ and $\gamma$ are two bi-orthogonal functions, then their Zak transforms are related by the equation:
\begin{equation}
\label{zak9}
Z_{g}^{*}(t,\omega) \cdot Z_{\gamma}(t,\omega) = \frac{1}{T},
\end{equation}
where $T$ is the period of the signal, or the window size. This means that Eq. (\ref{zak8}) can be rewritten:
\begin{equation}
\label{zak10}
a_{nm} = \text{ 2D~IDFT} \left ( T \cdot Z_{s} (t,\omega) \cdot Z_{\tilde g}^{*}(t,\omega) \right ).
\end{equation}
The striking consequence of (\ref{zak10}) is that in order to compute the PGBZ coefficients we no longer need to compute the overlap matrix and its inverse. We need only the Zak transform of a single function: the biorthogonal of the fiducial Gaussian. As a result, the Zak transform of a signal having $N_{t}N_{\omega} = N $ points is a matrix with $N$ elements, and we do not need to compute the huge $N^2$ overlap matrix.

In order to reconstruct the signal from its PGBZ map (\ref{zak10}) we just need to apply the following recipe (with the proper normalization constants): 
\begin{enumerate}
\item{Take the 2D DFT of the PBGZ.}\\
\item{Divide by the Zak transform of the fiducial Gaussian.}\\
\item{Take the inverse Zak transform of the result.}\\
\end{enumerate}
One might think from (\ref{zak10}) that we do not need to worry about the singularity of the Zak transform of the Gaussian since we multiply rather then divide by $Z_g^*$. However, the reconstruction of the signal from its PGBZ coefficients requires the dividing by $Z_g^*$ that we avoided in the forward direction. In other words we still have to bypass the singularity using the same recipe as before.

We now have in hand a very fast method (FFT order of computation) with very low memory requirement: for a 10 sec signal (sampled at 44.1 kHz) we only need 7 Mb (complex matrix of 665 $\times$ 665 in double precision). A 1 sec file now requires 0.7 Mb instead of 30 Gb. The computation time also becomes extremely short as we shall see in the next section.

\section{Results}
In order to evaluate the performance of the PGBZ method we compare it to two existing schemes, the Short-Time Fourier Transform (STFT) and the Discrete Wavelet Transform (DWT). Both schemes are well known and have been widely documented. 
The excellent paper by Zhivomirov \cite{Zhi} gives a good mathematical description of the STFT method. It also provides a state-of-the-art implementation of the algorithm as well as recommendations for choosing parameters for the method, including the analysis and synthesis windows. We followed the author’s recommendations, in particular those related to the invertibility conditions. In summary we adopt the following protocol:
\begin{itemize}
\item {Analysis window: Blackman-Harris. Synthesis window: Hamming.}
\item {Maximum hop size for perfect reconstruction = window length/8, which is equivalent to 80\% overlap (OLA condition). }
\item {Suitable zero-padding is performed at both extremities of the waveform in the STFT representation in order to obtain a correct reconstruction of the signal. }
\end{itemize}
% THE LAST TWO ITEMS AREN'T CLEAR TO ME
Additional tests were performed using other parameter sets for the STFT, e.g. changing the overlap to 50\%. See the Supplementary Material.

STFT schemes are widely used in audio analysis and processing. However, when signals are highly non-stationary (strong time-frequency correlation) the STFT leads to an inefficient TF representation. Wavelet transforms were developed to address this issue \cite{Ma99,Gr01}. The latter are both more flexible and more localized than Fourier representations, which makes them effective in signal and audio compression \cite{Rom,Kha,Man}. The choice of the wavelet family as well as the depth of the decomposition affects the quality of the compression process. We found that the Daubechies5 wavelet was most appropriate for our purpose;  the number of decomposition levels was varied between 5 and 10. We have tested other families, e.g. Symlet and Coiflet (see the Supplementary Material), but their performance is at best equal and generally inferior.
\subsection{The compression process}
Time-Frequency representation is the first step towards signal compression and provides the foundation for all further compression steps. Additional operations such as quantization and coding can be applied further, but since they operate in the same manner on the modified TF coefficient maps obtained by the three methods, they will not be considered in this study. 
Our implementation of the compression simply involves removing as many TF coefficients as possible, using some threshold criterion and reconstructing the signal from the modified TF map. In order to compress the signal, the coefficient map is divided into sub-spaces of sorted (ascending unique) values. At each compression step, one subspace is removed (i.e. set to 0). The compression process is continued until a given percentage of the initial number of TF coefficients is removed, and the resulting map is then inverted to reconstruct the compressed signal. Finally, the Mean Square Error (MSE) is calculated as a function of the remaining number of non-zero coefficients, where the MSE is defined as:
\begin{equation}
\label{res1}
\mathrm {MSE} (\%) = 100 \cdot \frac {\| s_{\mathrm {orig}}-s_{\mathrm {recons}}\|} {N [ {\mathrm {max}} (s_{\mathrm {orig}})-{\mathrm {min}}(s_{\mathrm {orig}})} .
\end{equation}
Here $s_{\mathrm {orig}}$ is the original signal, $s_{\mathrm {recons}}$ is the reconstructed signal using the compressed data, $N$ is the length of the signal and $||$  $||$ stands for the Euclidian norm. One can choose other ways to calculate the MSE, in particular by changing the normalization factor, but as long as the same formula is used for the three methods this choice has no real importance. We have also compared the three methods using metrics other than the MSE, for example psychoacoustic metrics. The qualitative conclusions remain the same; for details see the Supplementary Material. The Supplementary Material also contains links to several .wav files so the reader can compare the quality of the reconstruction for him/herself. 

The quality of the compression was also evaluated by (a) comparing the waveforms of the original and the maximally compressed signals, and (b) listening to the audio itself (i.e. the compressed sound). MSE is a good quantitative measure but it is an overall, average criterion, while waveform and audio can better reflect how close the compressed signal matches the original. 
The basic flowchart of the compression process is shown in Fig.  ~\ref{fig3}. It is the same for all methods.
\begin{figure}[!t]
\centering
\includegraphics[width=4in]{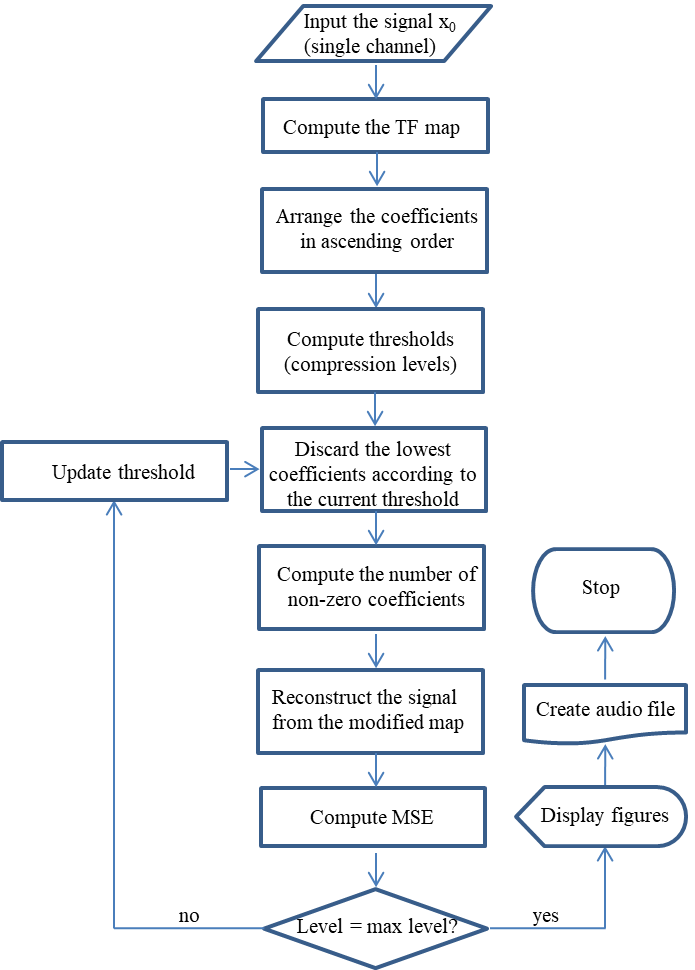}
\caption{Flowchart of the compression process.}
\label{fig3}
\end{figure}

\subsection{Reconstructing the signal}
As mentioned above, the PGBZ representation allows reconstructing the original signal from its PGBZ map.  Let us first consider a signal that consists of a simple sine in which a glitch has been inserted. Going from the signal to the PGBZ representation and back produces a reconstructed signal virtually indistinguishable from the original one. Although extremely small (10$^{-15}$), there is an error in the reconstructed signal with a periodic spiky shape, as one can see in Figure \ref{fig 4}. 
\begin{figure}
\centering
\subfloat[]{\includegraphics[width = 3.8in]{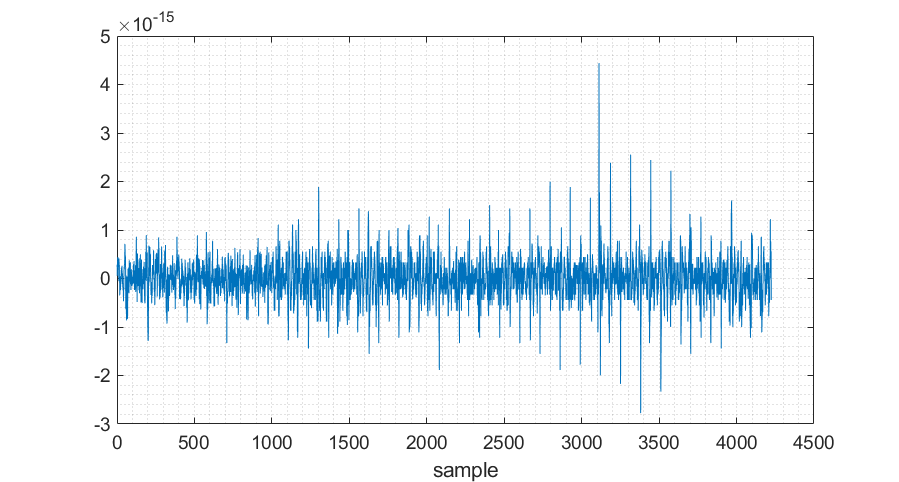}} \\
\subfloat[]{\includegraphics[width = 3.8in]{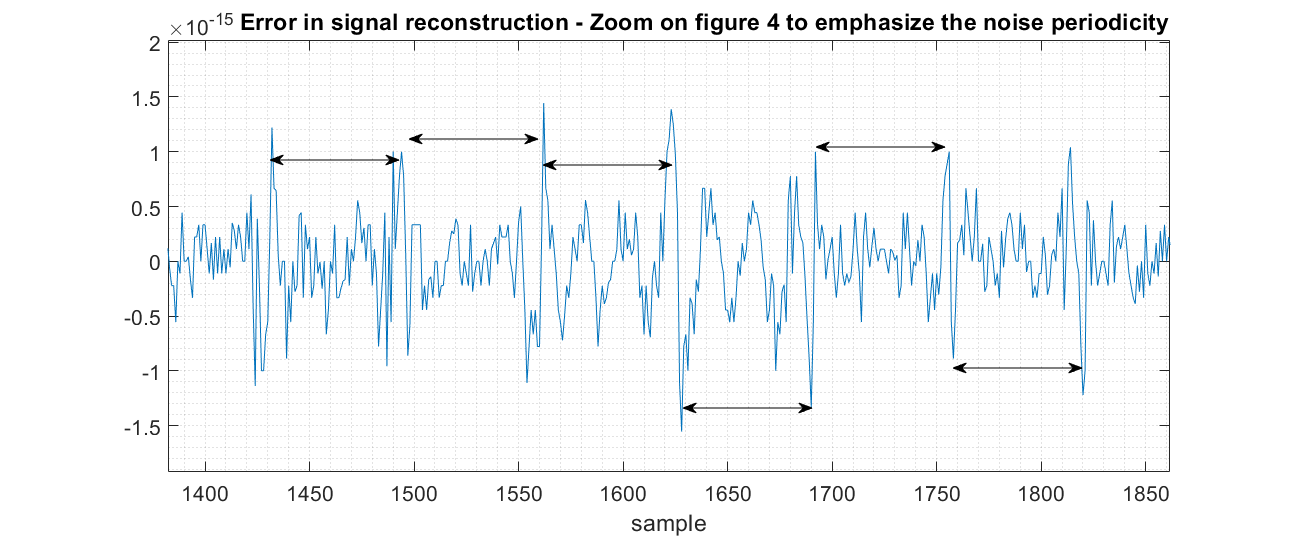}}
\caption{(a) The difference between the original and the PGBZ reconstructed signal (single sine + glitch).(b) Zoom on (a). The arrows all have the same length. Note the periodic, spiky shape of the error in the reconstructed signal.}
\label{fig 4}
\end{figure}
This pattern appears even when the signal is reconstructed using the full Gaussian basis and its biorthogonal counterpart, i.e. using the PGB formalism without Zak transform. At the end of the compression process 96\% of the coefficients have been set to zero. At this stage the amplitude of the periodic pattern has considerably increased and the resulting noise cannot be ignored anymore, as can be seen in Figure ~\ref{fig5} (orange curve). When analyzed closely, it appears that the shape of the pattern is very similar to the shape of the functions biorthogonal to the Gaussians. Careful checks show that numerical error is not responsible for the pattern, which seems to be intrinsic to the way we build the compressed signal from the remaining basis functions.
%IN THE FOLLOWING PARAGRAPH, BE CONSISTENT WITH EARLIER NOTATION ABOUT BASIS SETS AND SUBSCRIPTS (BF, MATHBF, WHERE TO PUT BRACKETS, ETC.)
In the early nineties, Genossar and Porat published an elegant algorithm that considerably improves the accuracy of the reconstruction \cite{Gen}, and almost completely removes the spiky artifact. Consider a signal $s(t)$ expanded on a periodic Gabor basis of size $N$, $\bf{G}_{N}$. After compression only $K$ coefficients remain ($K \ll N$), i.e. only $K$ biorthogonal basis vectors contribute to the expansion of the compressed signal. This reduced biorthogonal basis is denoted  $\bf {\hat {\Gamma}}_{K}$ . 
Following Porat we define the corresponding reduced Gabor basis set 
$\bf { \hat {G} _{K}} $ by $\bf { \hat {G}_{K} = \hat {\Gamma}_{K}  (S_{K}^{ \hat {\Gamma}}) ^{-1} }  $ 
where $ \bf {S_{K}^{ \hat {\Gamma}} 
= \hat {\Gamma}_{K}^{ \dag } \hat {\Gamma}_K }$,
(the $\bf{S}$ matrix being now defined in terms of the contracted set only). The reconstructed signal $s_{r} (t)$ is given by:
\begin{equation}
\label{res2}
s_{r}(t) = \sum_{i = 1}^{K} \hat {c}_{i} \hat {\gamma} _{i},
\end{equation}
where  $\hat {c}_{i} = \braket{\hat {g} _{i} | s },$  $\hat {g} _{i} \in   \bf {\hat {G}_K },$ $\hat {\gamma} _{i} \in   \bf {\hat {\Gamma}_K }.$
According to \cite{Gen}, $s_r$  is the closest to $s$ in the minimal norm sense, from among the vectors of the space spanned by $\bf{\hat {\Gamma}_K}$. Applying Porat’s scheme to our signal produces the magenta curve shown in Fig.  ~\ref{fig5}. The spiky pattern has been almost entirely eliminated. Unfortunately, Porat’s scheme requires the computation of the full biorthogonal basis $\bf {\Gamma_{N}^{G}}$, including the overlap matrix $ \bf {S_N^G}$  and its inverse, limiting the PGB method to very small signals.
\begin{figure}
\subfloat[]{\includegraphics[width = 4.2in]{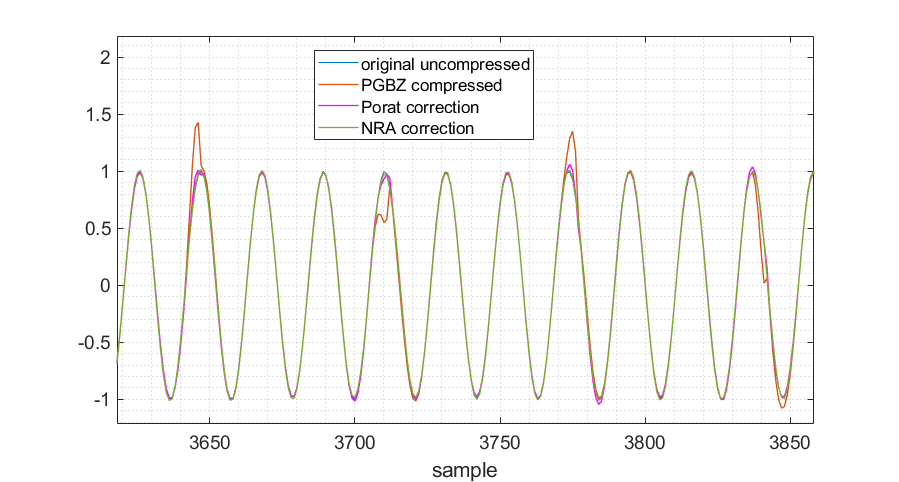}} \\
\subfloat[]{\includegraphics[width = 4.2in]{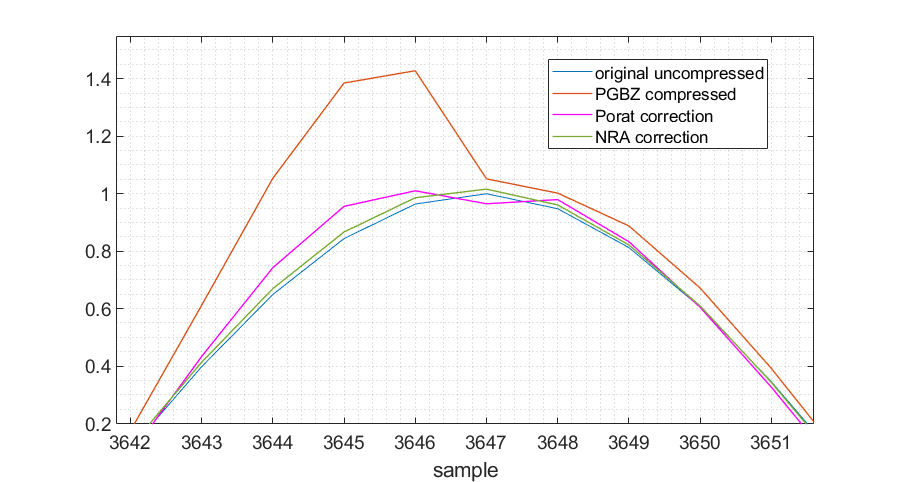}}
\caption{Original signal vs. compressed and corrected signals (5a top). Zoom around sample 3646 (5b bottom)}
\label{fig5}
\end{figure}
Is there a less expensive way to implement the Porat algorithm? We found that a spectral filter of the type used in commercial audio editing software like e.g. the Audacity package \cite{Aud} provides a semi quantitative approximation to the Porat correction. After learning the spectral characteristics of the artifact, the filter truncates spurious frequencies when they get larger than a certain threshold that depends on the frequency band. The green curve in Fig. ~\ref{fig5} shows the compressed PGB filtered by the Noise Reduction Algorithm (NRA). 
\begin{figure}[!t]
\centering
\includegraphics[width=4.2in]{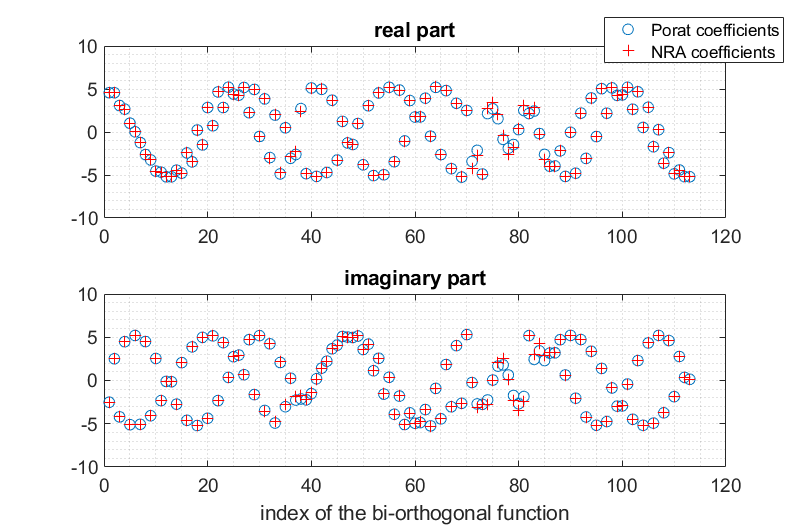}
\caption{Real and imaginary values of the Porat and NRA expansion coefficients (sine signal).}
\label{fig6}
\end{figure}
Fig.  ~\ref{fig6} compares the distribution of the coefficients $\hat {c} _i$ obtained with the Porat algorithm with the expansion coefficients of the PGB signal 
$\hat {d}_{i} = \braket{\hat {g} _{i} | s_{i}} $ after passing through the NRA filter. These two sets of coefficients (real and imaginary parts), are plotted as a function of the index of their corresponding $\bf {\hat {\Gamma}_K}$  functions. It is striking that the distributions of the two sets are almost identical: the “ad hoc” filter has produced a signal almost identical to the optimal approximation of the compressed data. 

In order to confirm this non-trivial finding we have performed the same comparison on a more complex audio signal extracted from a guitar recording. The resulting waveforms are shown in Fig.  ~\ref{fig7}. Note the very spiky pattern on the compressed PGBZ reconstructed signal.
\begin{figure}
\subfloat[]{\includegraphics[width = 3.8in]{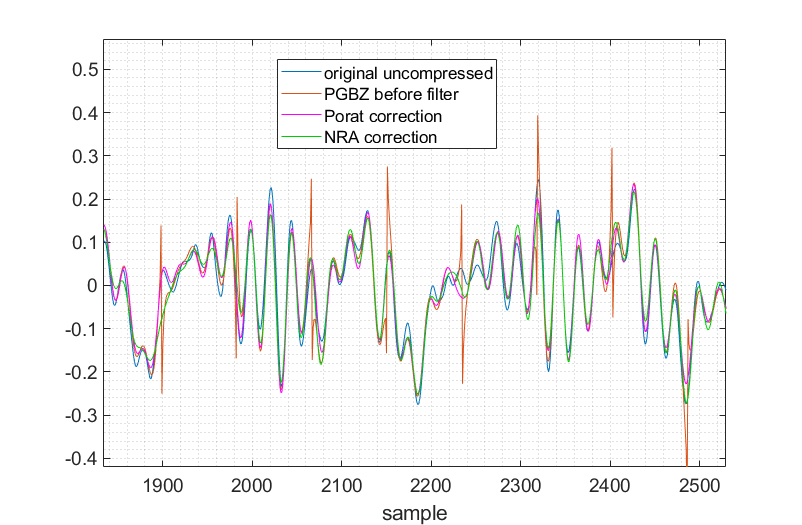}} \\
\subfloat[]{\includegraphics[width = 3.8in]{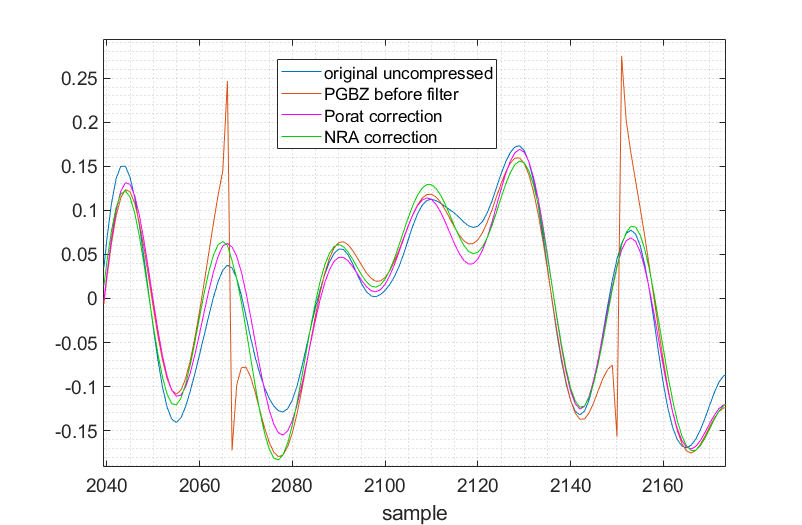}}
\caption{Waveforms of audio signal before and after 96\% compression. The bottom plot is a zoom of the top one}
\label{fig7}
\end{figure}
The  $\hat {c} _i$   and $\hat {d} _i$  distributions are shown in Fig.  ~\ref{fig8}. We can see again that the “optimal” coefficients of the reduced basis set $\bf { \hat {\Gamma}_K } $  are almost the same for the Porat and filtered PGBZ. 
\begin{figure}[!t]
\centering
\includegraphics[width=4in]{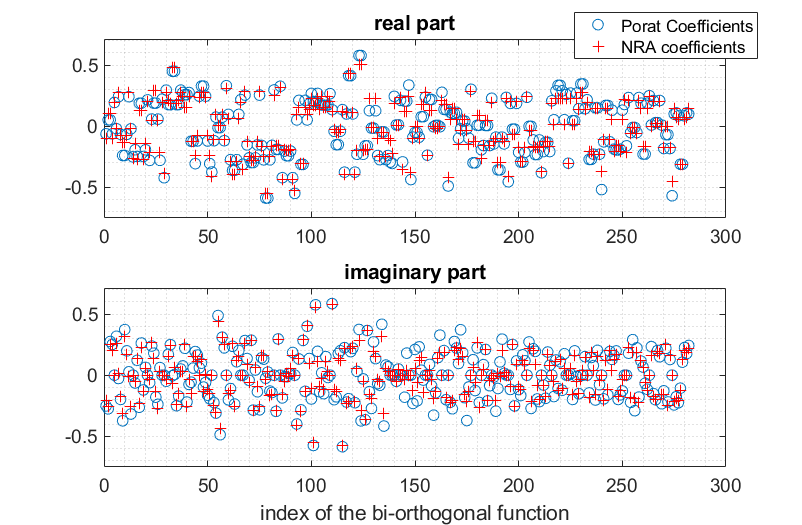}
\caption{Real and imaginary values of the Porat and NRA expansion coefficients (audio signal).}
\label{fig8}
\end{figure}
Due to the huge size of the overlap matrix we cannot perform a comparison of the two methods for large signals. However, based on results for smaller signals, we believe that applying the noise reduction algorithm to PGBZ for large reconstructed signals will be very similar to that obtained with Porat’s scheme if it were feasible to compute the latter. Therefore from this point on, all the PGBZ results presented in this paper were passed through the NRA described above.

\subsection{Audio files}
The comparison of PGBZ with STFT and DWT was performed on 10 audio files including 4 instrumental samples (piano, drum, flute and guitar), 2 vocal samples (bossa-nova, acapella), 2.5 minutes from the 6th Symphony of Beethoven and 4 speech samples. The music samples come from the SampleFocus site \cite{foc}, and the speech files from the CMU-ARCTIC speech synthesis databases from the Carnegie-Mellon University \cite{Kom}, and the OpenSLR dataset \cite{OSLR}. In all PGBZ computations the window size has been taken to be the integer part of the square root of the signal length. No zero-padding was necessary. The maximum level of compression was limited to 96\% i.e. at the end of the compression process only 4\% of the TF map coefficients are used to reconstruct the signal. Beyond this limit the audio signal significantly deteriorates. Fig.  ~\ref{fig9} shows the MSE in \% as defined in (\ref{res1}), as a function of the remaining number of non-zero coefficients.
\begin{figure}
\subfloat[]{\includegraphics[width = 3in]{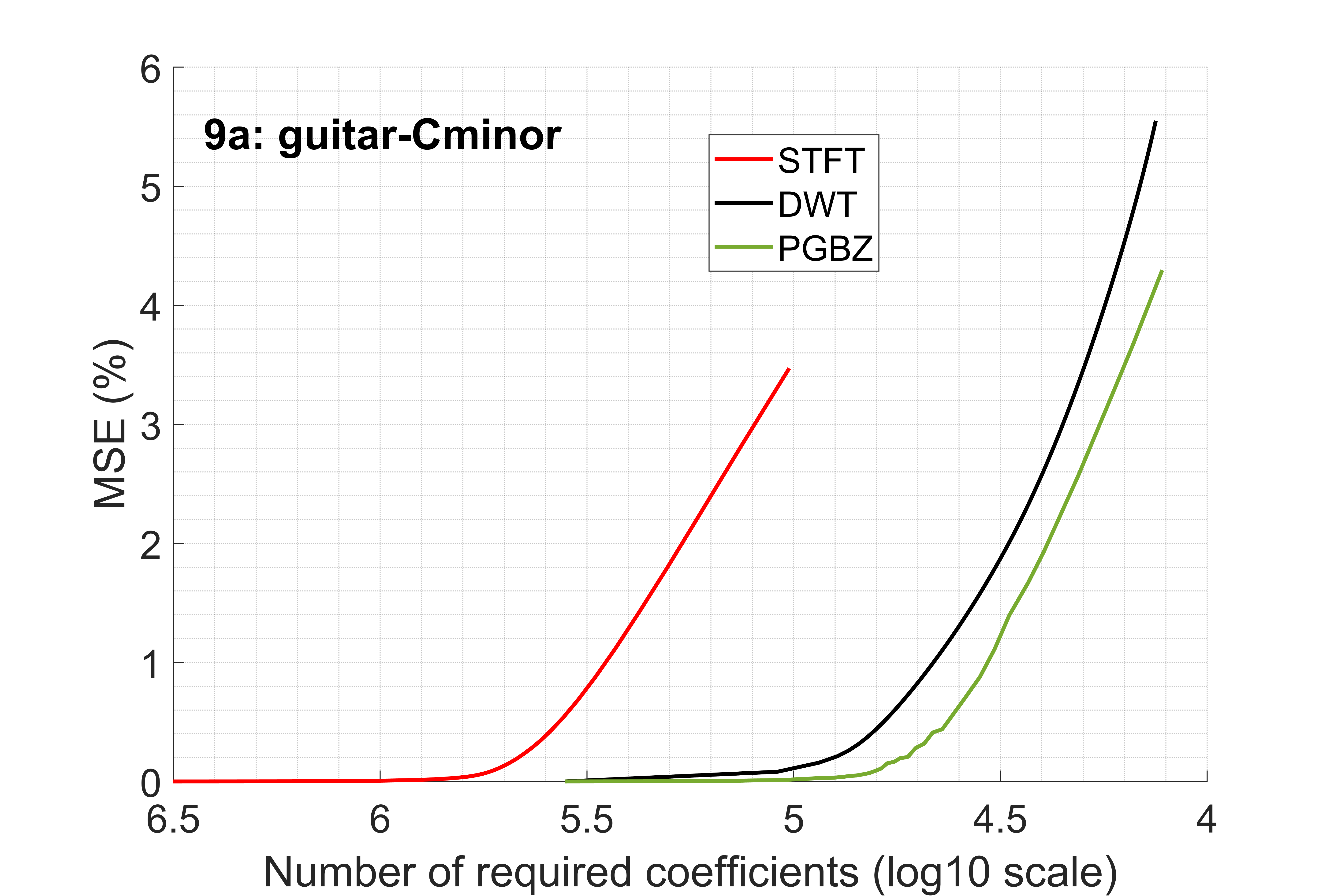}} \\
\subfloat[]{\includegraphics[width = 3in]{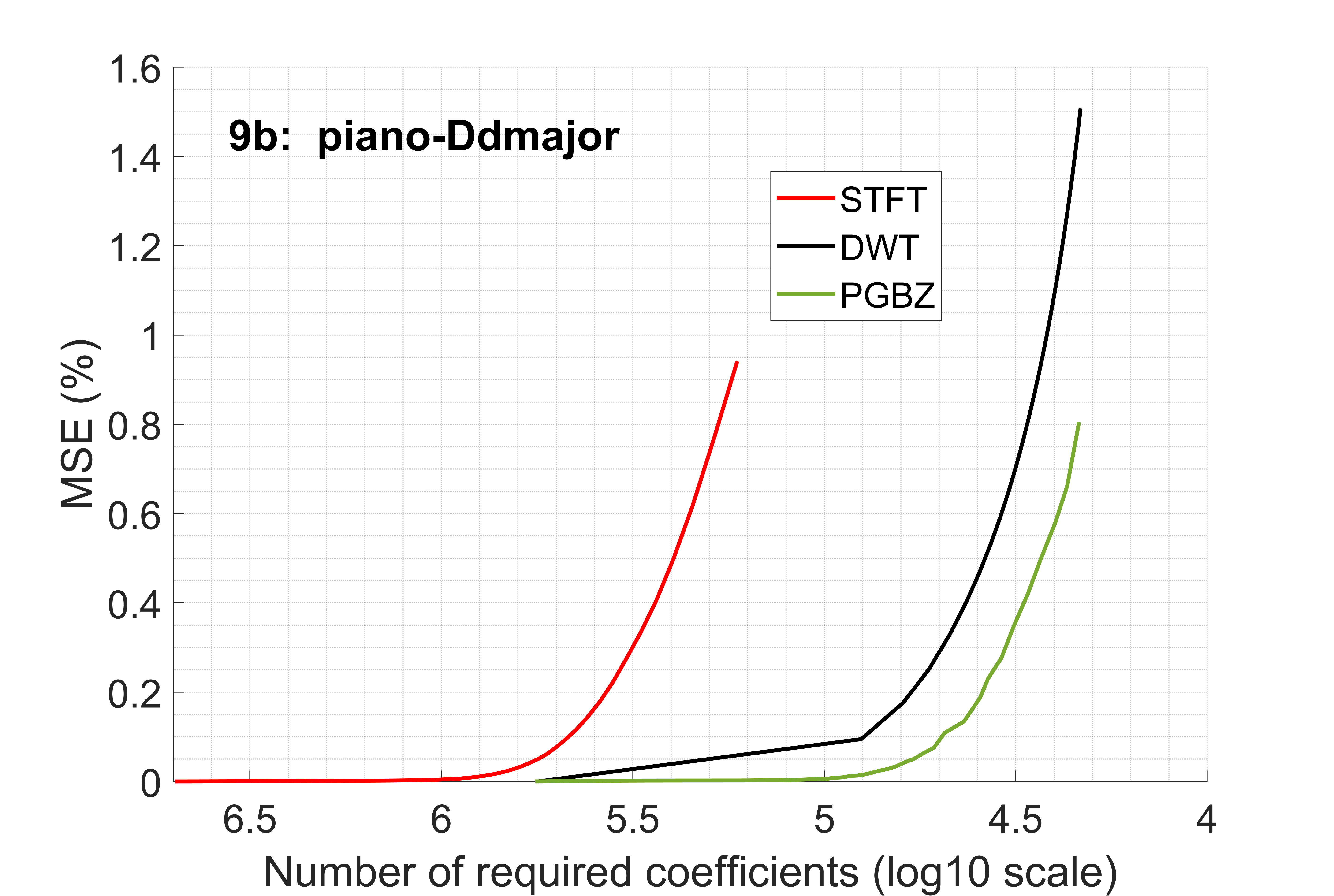}}\\
\subfloat[]{\includegraphics[width = 3in]{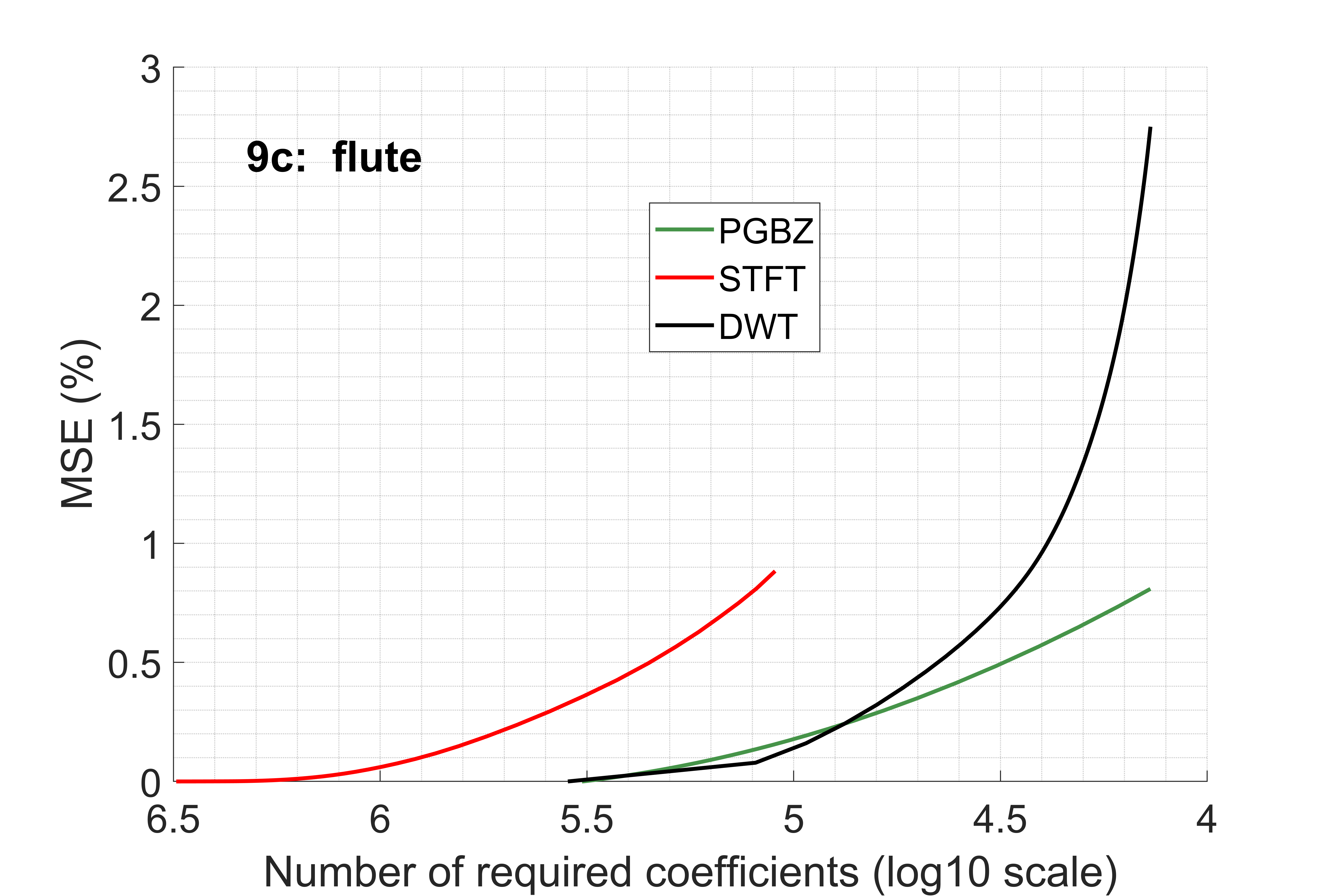}}\\
\subfloat[]{\includegraphics[width = 3in]{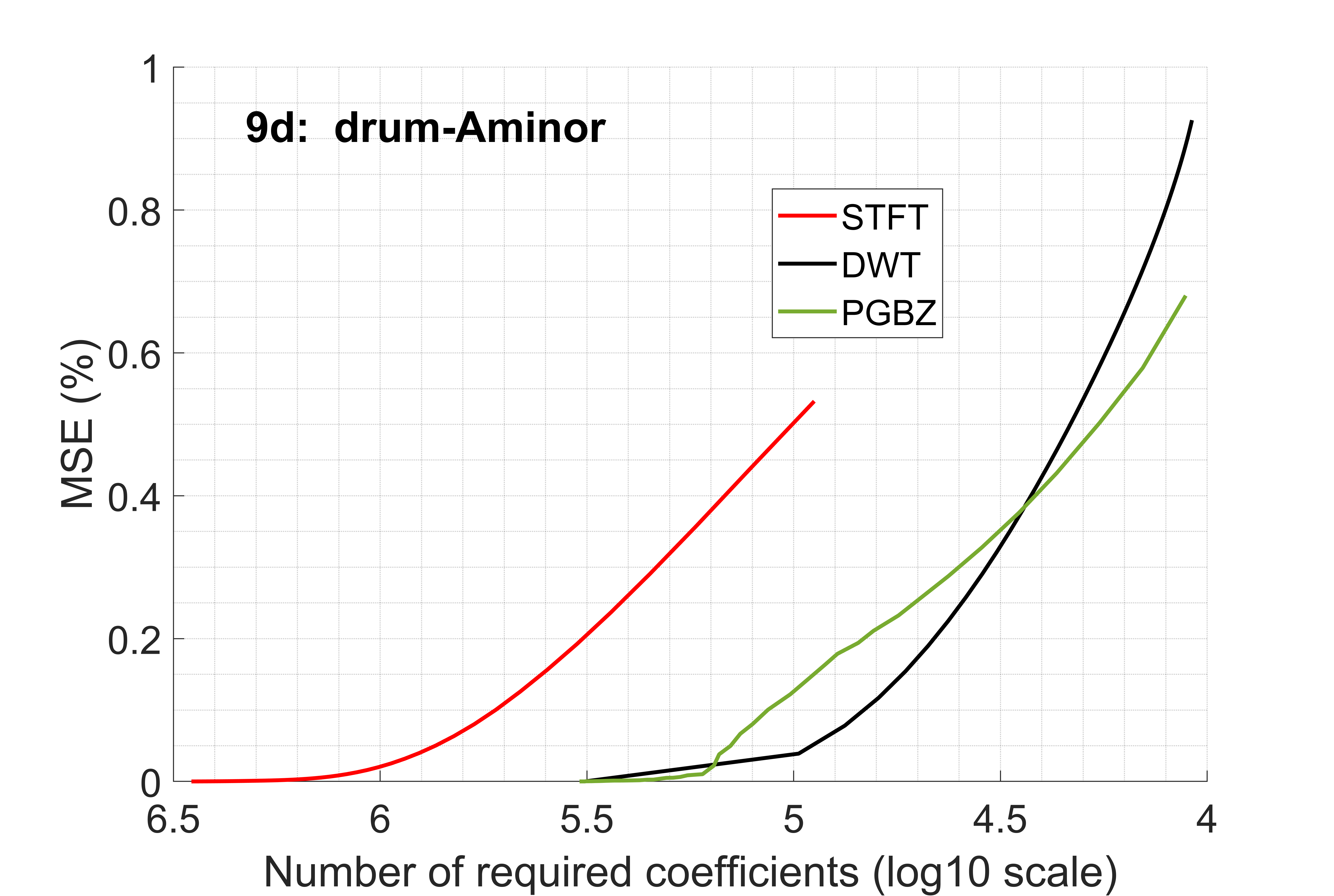}}
\end{figure}
\captionsetup[subfigure]{labelformat=empty}
\begin{figure}
\subfloat[(e)]{\includegraphics[width = 3in]{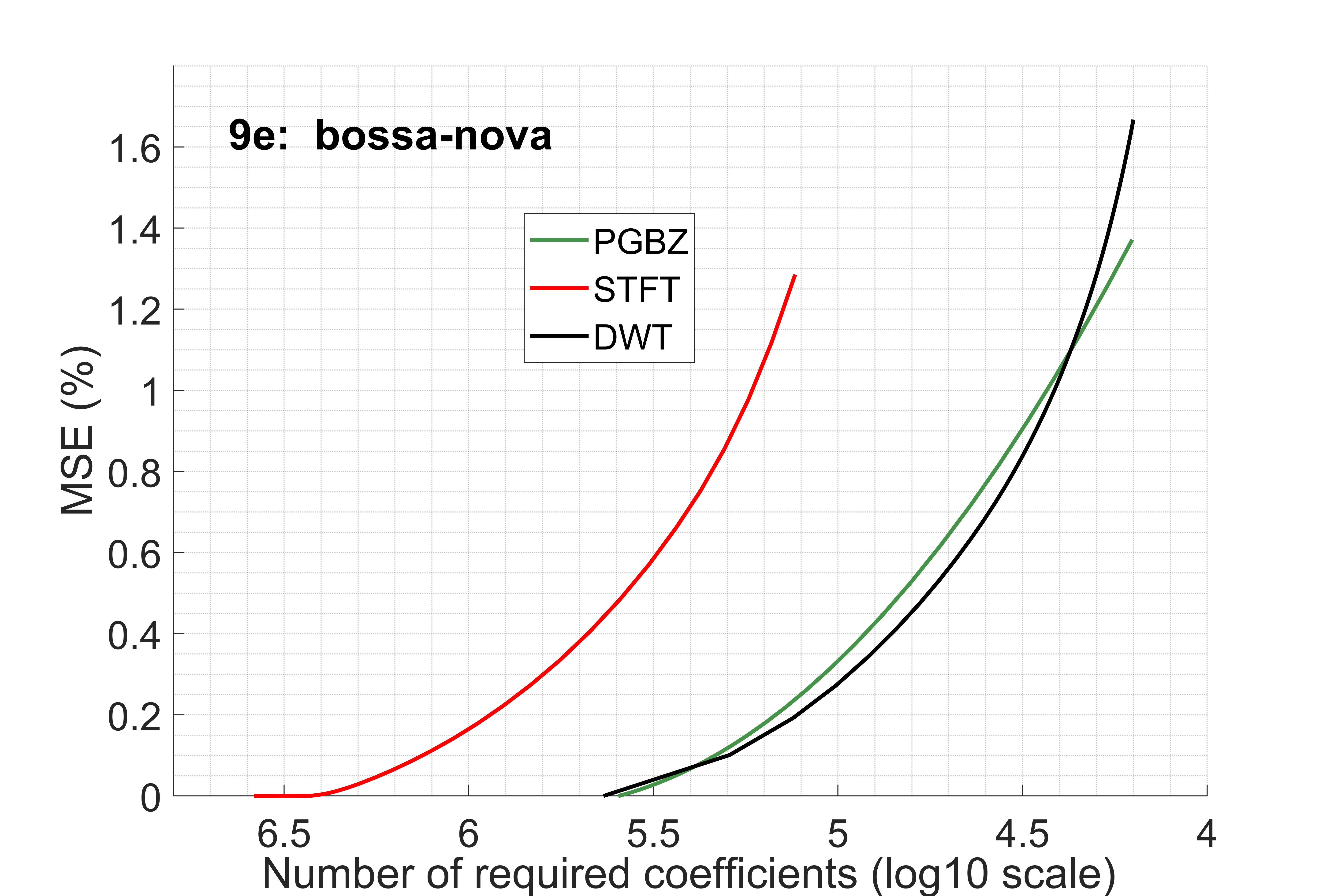}}\\
\subfloat[(f)]{\includegraphics[width = 3in]{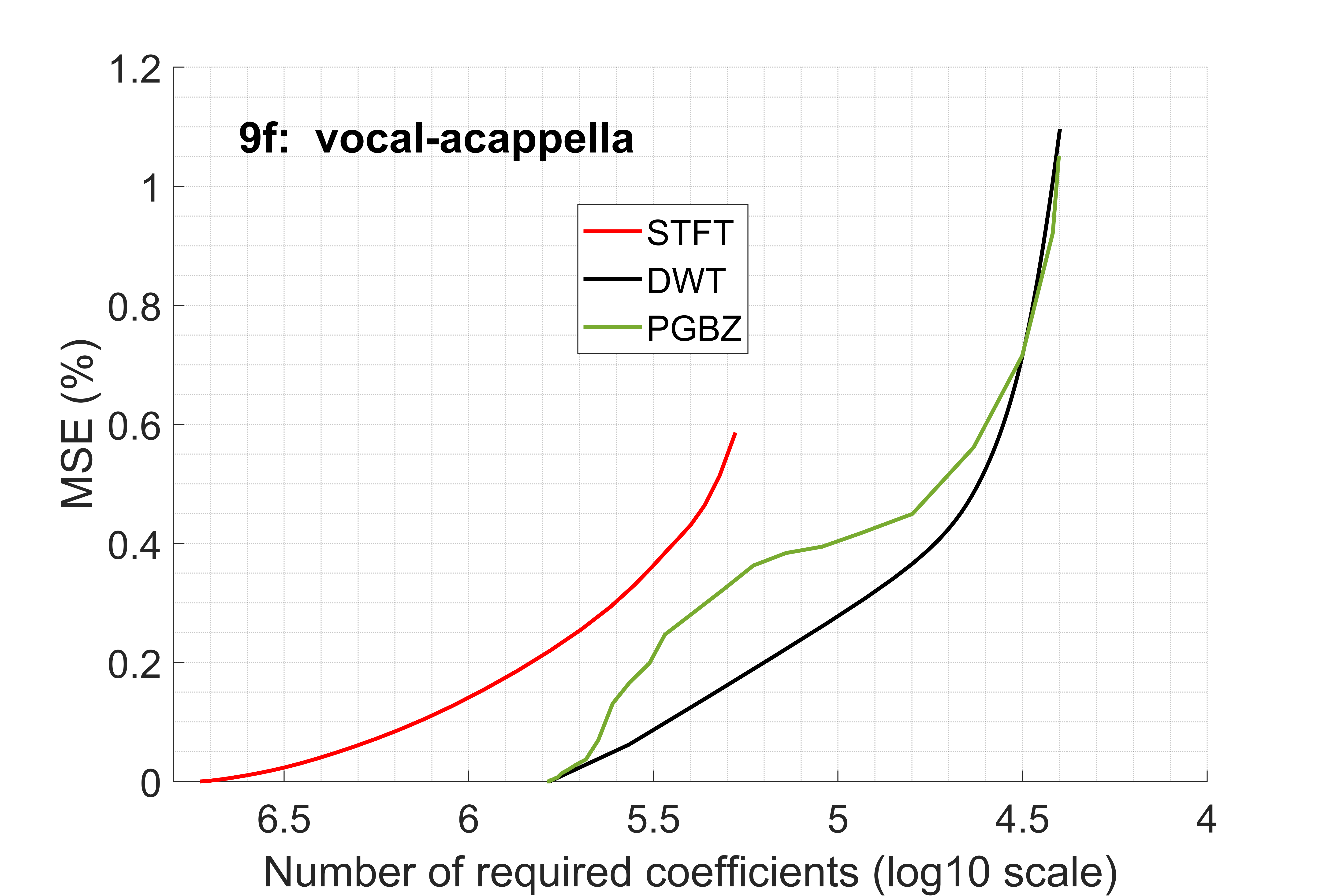}}
\caption{MSE vs. Number of coefficients used for reconstruction (in log scale), for 6 music samples. Note that the x-axis is plotted in inverse order, starting from the full map size on the left to the compressed map on the right.}
\label{fig9}
\end{figure}
\captionsetup[subfigure]{labelformat=parens}
In all cases, the STFT is outperformed by the two other methods by at least an order of magnitude. This is true even at zero compression, and it is mostly due to the large overlap between windows required by the OLA condition in the STFT. Note that DWT and PGBZ start at the same full coefficient map size (at zero compression both representations have the same dimension). 
As the number of non-zero coefficients decreases, DWT and PGBZ start diverging. For the guitar and piano PGBZ performs better than DWT at all compression levels. In these cases the PGBZ error stays close to zero even when the number of relevant coefficients decreases by an additional factor of 3 relative to DWT; only from this point on does the error starts growing. The flute and bossa nova are intermediate cases where the two methods are almost equivalent until some level at which the DWT falls behind the PGBZ. In the drum sample, DWT performs better than PGBZ until the curves cross and PGBZ performs better. The a cappella case is the only case where DWT is always “cheaper” than PGBZ, although at high compression levels the two methods exhibit the same MSE values. Also, except for this example, DWT exhibits a steeper slope and terminates with an error much larger than PGBZ.
Fig.  ~\ref{fig10} shows the performance of the three methods for speech samples. In all cases PGBZ performs as well as or better than DWT. 
\begin{figure}
\subfloat[]{\includegraphics[width = 3in]{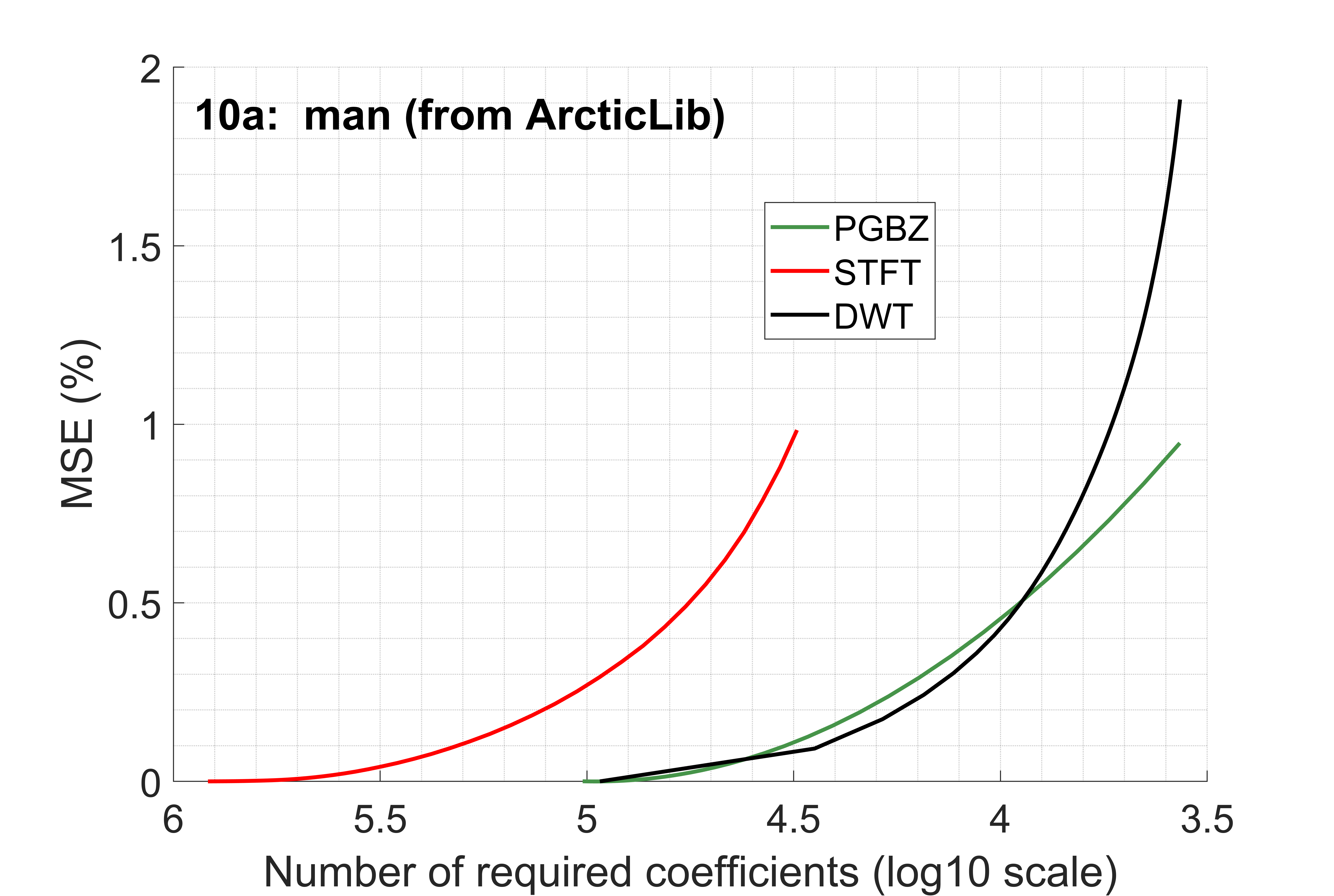}}\\ 
\subfloat[]{\includegraphics[width = 3in]{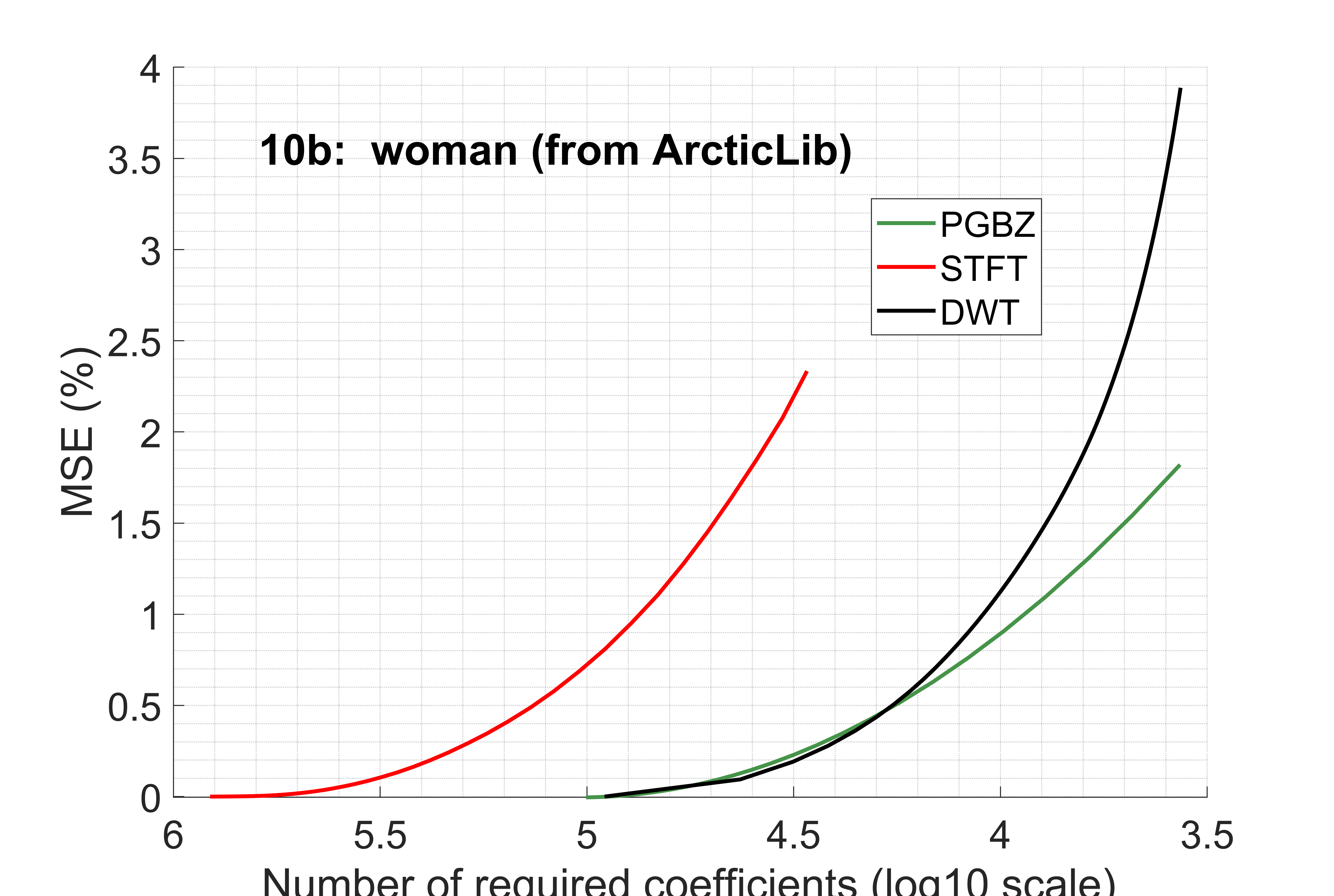}}\\
\subfloat[]{\includegraphics[width = 3in]{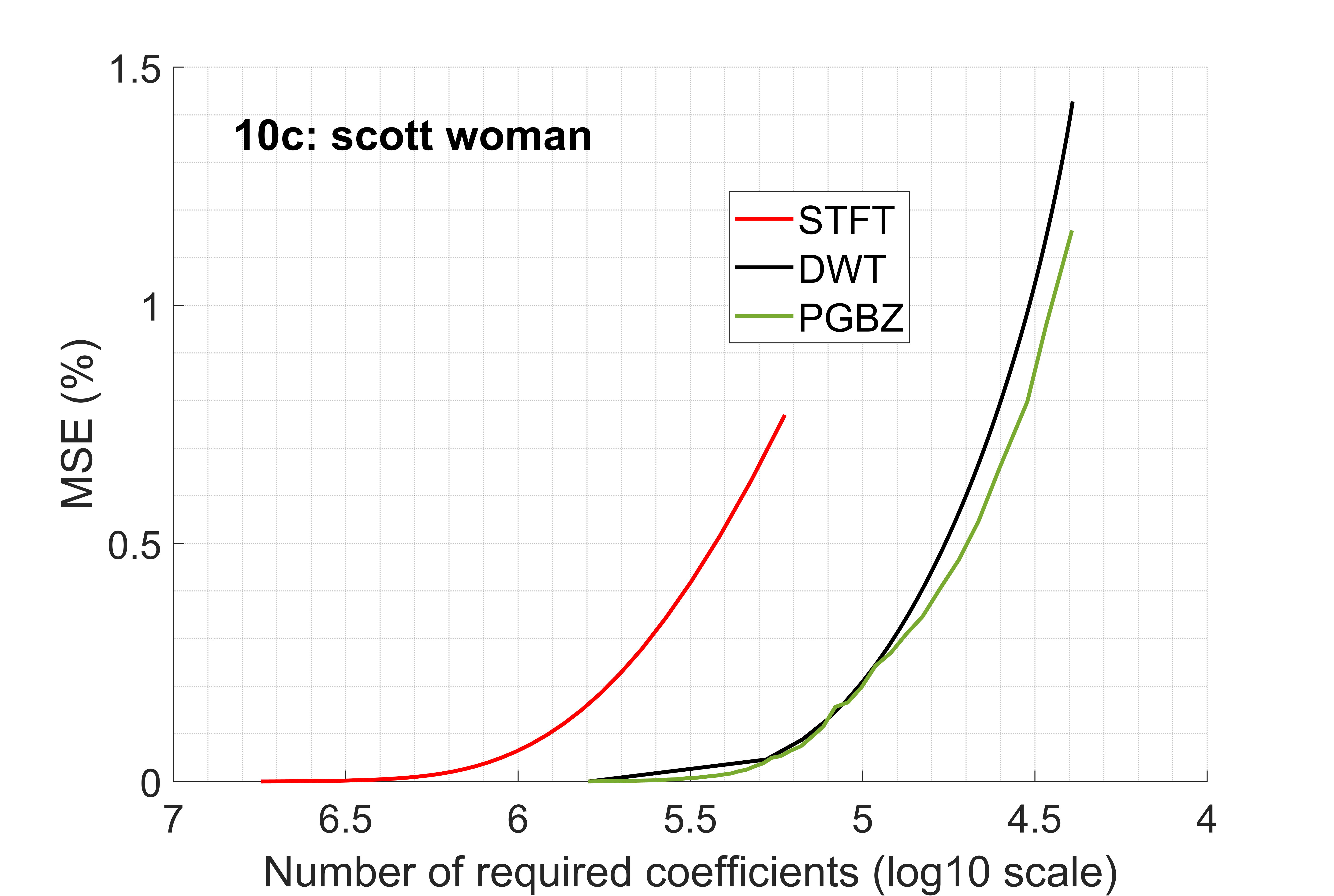}}\\
\subfloat[]{\includegraphics[width = 3in]{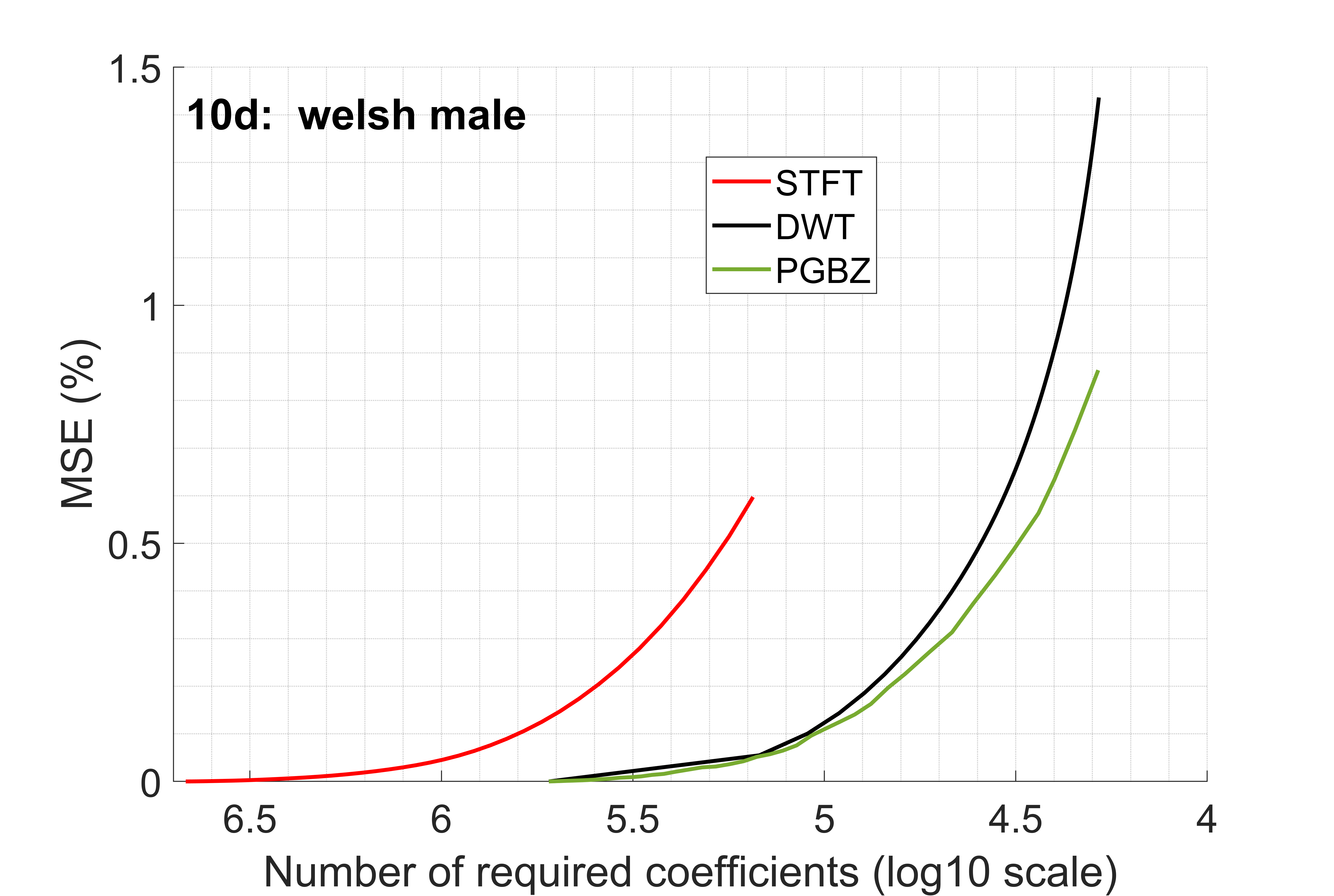}}
\caption{MSE vs. Number of coefficients used for reconstruction (in log scale), for 4 speeches samples.}
\label{fig10}
\end{figure}

All the audio files tested above were short samples (8 to 10 sec). In order to test the run time performance we now consider a much longer music sample, the first 2.5 minutes of 
the 6th Symphony of Beethoven. The MSEs of the three methods are shown in Fig \ref{fig11}. 
\begin{figure}[!t]
\centering
\includegraphics[width=3.8in]{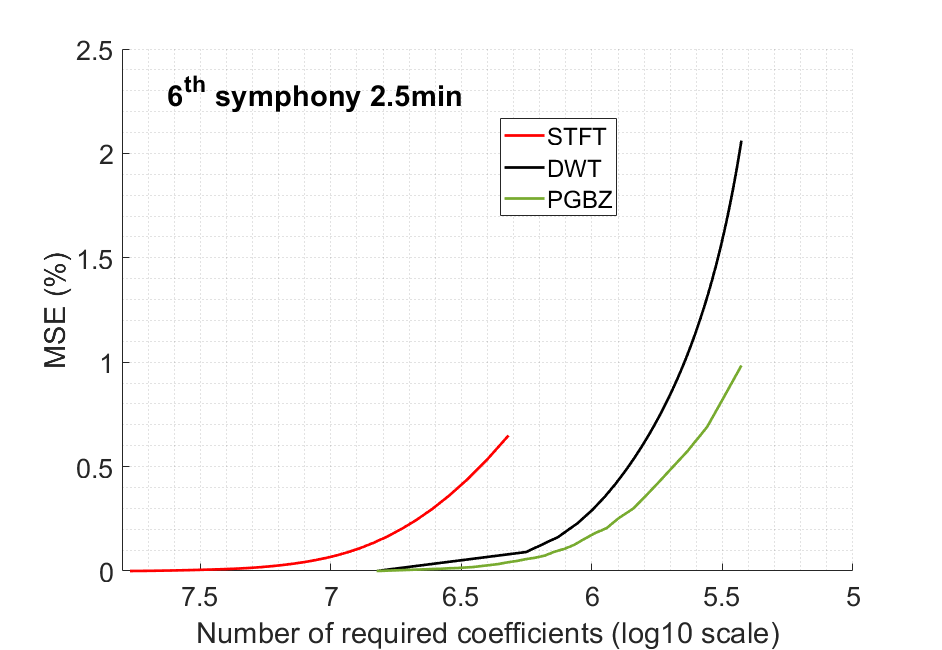}
\caption{MSE vs. Number of coefficients used for reconstruction (in log scale) for the first 2.5 minutes of the 6th Symphony of Beethoven.}
\label{fig11}
\end{figure}

The computations were performed on an Intel i7-10850H processor. Results are shown in Table I.  NMSE stands for Normalized MSE: it is the product of the three variables at the last level of compression --- the MSE, the minimum number of remaining non-zero coefficients (denoted $K$) and the CPU time required to reach this level. We see that although DWT is about 15\% faster than PGBZ, the best final score by far is reached by the PGBZ method.
\begin{table}
\begin{center}
\caption{Comparison ot the run times and normalized errors.}
\label{tab1}
\begin{tabular}{| c | c | c |c|}
%SHOULD ADD LINES FOR MSE AND MINIMUM NUMBER OF REMAINING NON-ZERO COEFFICIENTS
\hline
Method & STFT & DWT & PGBZ\\
\hline
MSE & 0.65 & 2.06 & 0.98 \\
\hline
log($K$) & 6.3 & 5.43 & 5.43\\  
\hline
CPU (sec) & 232 & 52 & 64\\
\hline
Maximum NMSE & 980 & 560 & 340 \\ 
\hline
\end{tabular}
\end{center}
\end{table}

It is also interesting to compare the actual waveforms of the compressed signals. In order to make a fair comparison we will compare the PGBZ result with the STFT and the DWT separately. For the former comparison we plot the waveform of the STFT at the maximum error and the PGBZ signal at the same error. For the latter comparison we simply plot the waveforms at the maximum level of compression. Both comparisons also include the original, uncompressed signal. The two plots are shown in Figs.  ~\ref{fig12} and  ~\ref{fig13} respectively. 
As one can see from Fig.  ~\ref{fig11}, despite their MSEs being similar, the PGBZ follows the original signal much more closely than does the STFT. Look in particular between 1.57 and 1.573: the STFT result tends to smooth the original waveform while PGBZ follows the uncompressed waveform. Converted into audio the STFT will sound both attenuated and lower (more bass) than the original (and the PGBZ) music.
\begin{figure}[!t]
\centering
\includegraphics[width=3.8in]{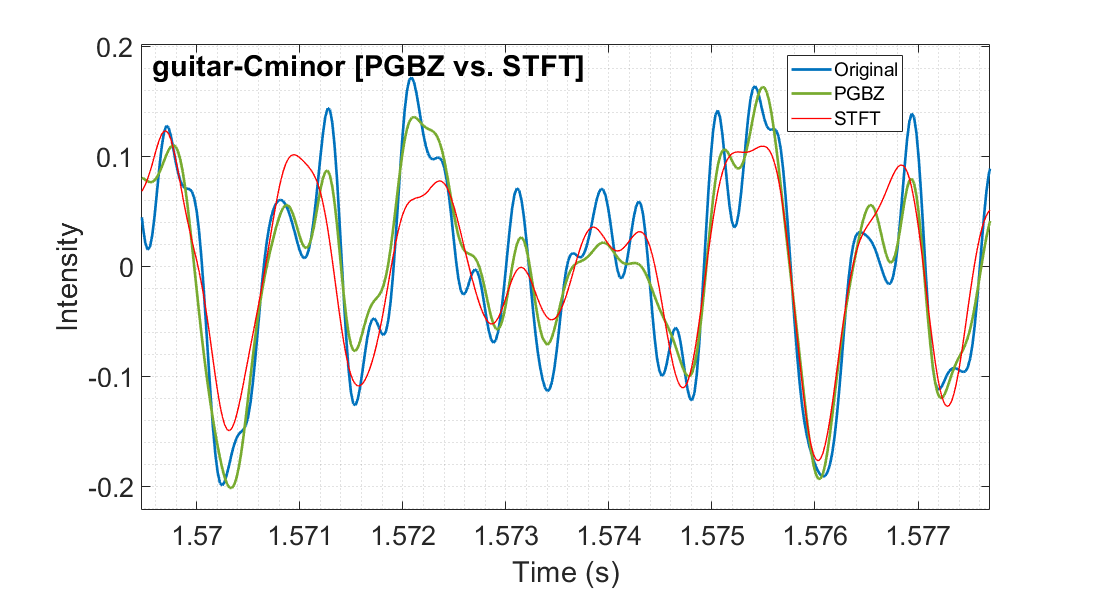}
\caption{Original vs. reconstructed signals (STFT and PGBZ) at the maximum error value reached by the STFT method.}
\label{fig12}
\end{figure}
\begin{figure}[!t]
\centering
\includegraphics[width=3.8in]{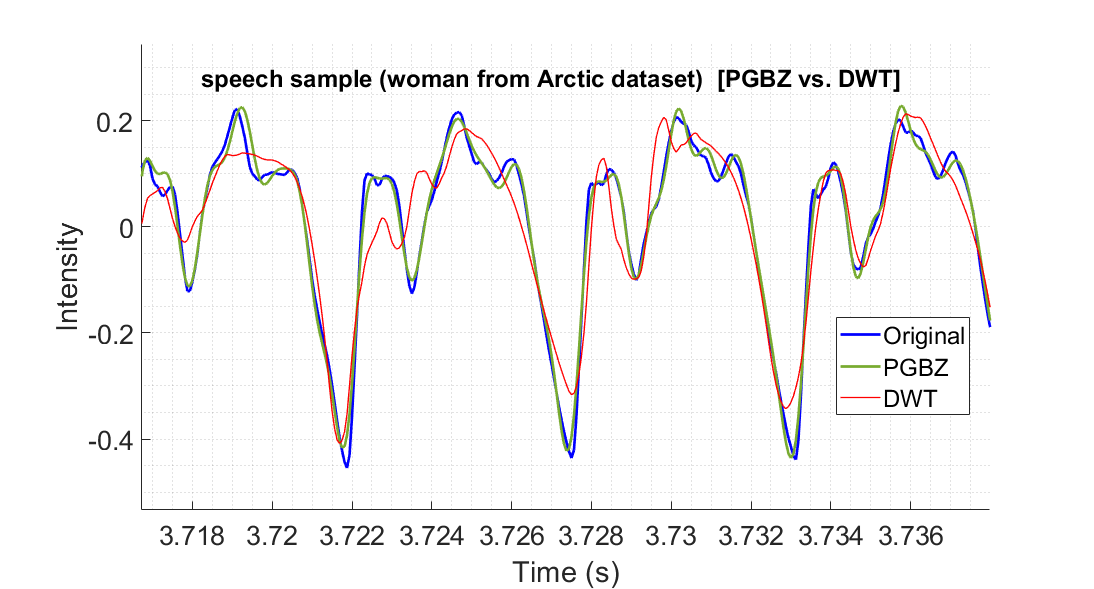}
\caption{Original vs. reconstructed signals (DWT and PGBZ) at the maximum compression level.}
\label{fig13}
\end{figure}
Turning to comparison with DWT, the difference in the global MSE is well reflected in the waveforms: the PGBZ result is much more accurate than the DWT. This can be heard also in the audio sounds produced from the waveforms. The phrase being pronounced can barely be recognized in the DWT compressed signal, while the PGBZ is almost indistinguishable from the original.

\section{Conclusion}
We have developed a new formalism for a stable and accurate Time Frequency (TF) representation of signals based on a Periodic Gabor representation combined with Biorthogonal exchange (PGB). The scheme becomes computationally very efficient when the coefficients are computed by Zak Transform, avoiding the need to compute the large overlap matrix that was necessary in a previous version of this work \cite{Shim3}. The method, called PGBZ, has been successfully used here to compress audio signals, including music and speech samples. Due to the strong locality of the TF coefficients, it performs much better than STFT and in many cases outperforms Wavelet Compression. The scheme can be easily extended to image compression and may offer an attractive alternative to state-of-the art methods existing in the literature.

The method has distinct advantages over the widely used STFT and DWT methods.
\begin{itemize}
\item The method is more efficient than STFT by an order of magnitude and competitive with or more efficient than DWT, for similar computational complexity, speed and memory.
\item The theory and implementation are simpler than DWT, and as opposed to DWT there are essentially no free parameters. 
\item The formalism is somewhat more complicated than STFT, but on the other hand STFT has multiple free parameters that must be optimized (e.g. the overlap factor), while again PGBZ has essentially no free parameters. 
\item As opposed to STFT and some implementations of DWT, PGBZ is a basis, not a frame, i.e. there is no overcompleteness in PGBZ.  This provides a certain uniqueness to the formalism and in principle provides an optimal representation since there is no overcompleteness. This is also the underlying reason why there are essentially no free parameters in PGBZ.
\item The method is faithful to Gabor’s original vision, in that it involves critical sampling of the time-frequency space using translated replicas of a fiducial basis function. However, Gabor’s original proposal is unstable (\cite{Daub,Balian,Low}), and even after fixing the problem with stability (\cite{Chin}), sparsification is highly inefficient (\cite{Shim1}) and a large matrix inversion is required. By combining the Fast Zak Transform with periodicity and biorthogonal exchange, we have here a method that uses identical replicas of a fiducial basis function to cover the time-frequency space that is stable, sparse and fast.
\end{itemize}
A disadvantage of the method is that for optimal performance of PGBZ, noise filtering should be done after compression. Admittedly this adds a layer of complexity to the implementation but led us to an unexpected and fascinating connection between commercial noise reduction software and the rigorous but much more expensive Porat procedure for reorthogonalizing the biorthogonal basis after compression. 

Clearly, further tests are required to assess the competitiveness of the PGBZ method. But we believe that the current work establishes it as a serious competitor to the widely used STFT and DWT methods, and as such the method is worthy of serious consideration, both from a practical and theoretical point of view. The new approach may turn out to be the method of choice for certain classes of applications, and additional variants of the method may be discovered that continue to improve its performance.

\section {Acknowlegments}
Financial support for this work came from the Israel Science Foundation (1094/16 and 1404/21), the German-Israeli Foundation for Scientific Research and Development (GIF), the Ben May Center for Theory and Computation as well as the historic generosity of the Harold Perlman family.\\

Helpful discussions with Profs. Yonina Eldar and Joshua Zak are gratefully acknowledged.

\end{document}